\pdfoutput=1
\documentclass[amsmath,amssymb,aps,prd,preprint,groupedaddress]{revtex4-2}
\usepackage{hyperref}
\usepackage{color}
\usepackage[table]{xcolor}
\usepackage{graphicx}

\usepackage{verbatim}
\usepackage{subfig}
\usepackage{url}
\usepackage{bbold}
\usepackage{slashed}

\usepackage{multirow}
\usepackage{threeparttable}
\usepackage{paralist}

\usepackage{natbib}
\usepackage{xspace}

\newcommand{\be}{\begin{equation}}
\newcommand{\ee}{\end{equation}}

\newcommand*{\MGMCatNLO}{\textsc{MadGraph5}\_aMC@NLO\xspace}
\newcommand*{\MadGraph}{\MGMCatNLO}
\newcommand*{\MADSPIN}{\textsc{MadSpin}\xspace}
\newcommand*{\MadSpin}{\MADSPIN}
\newcommand*{\PYTHIA}{\textsc{Pythia}\xspace}
\newcommand*{\Pythia}{\PYTHIA}
\newcommand*{\WpWmVBS}{\ensuremath{W^\pm W^\mp jj}}
\newcommand*{\ssWWVBS}{\ensuremath{W^\pm W^\pm jj}}
\newcommand*{\WZVBS}{\ensuremath{WZ jj}}
\newcommand*{\ttbar}{\ensuremath{t\bar{t}}}

\begin{document}
\title{Sensitivity to longitudinal vector boson scattering in semileptonic final states at the HL-LHC}

\author{Jennifer Roloff}
\author{Viviana Cavaliere}
\author{Marc-Andr\'e Pleier}
\affiliation{Physics Department, Brookhaven National Laboratory, Upton, New York 11973-5000, USA}
\author{Lailin Xu}
\affiliation{Department of Modern Physics and State Key Laboratory of Particle Detection and Electronics, University of Science and Technology of China, Hefei 230026, China}

\begin{abstract}
Longitudinal vector boson scattering provides an important probe of electroweak
symmetry breaking, bringing sensitivity to physics beyond the Standard Model as
well as constraining properties of the Higgs boson. It is a difficult process to
study due to the small production cross section and challenging separation of
the different polarization states. We study the sensitivity to longitudinal
$WV$ vector boson scattering at the high-luminosity Large Hadron Collider
in semileptonic final states. While these are characterized by larger background 
contributions compared to fully leptonic final states, they benefit from a higher
signal cross section due to the enhanced branching fraction.
We determine the polarization through full
reconstruction of the event kinematics using the $W$ boson mass constraint and
through the use of jet substructure. We show that with these techniques
sensitivities around three standard deviations at the HL-LHC are achievable,
which makes this channel competitive with its fully leptonic counterparts.
\end{abstract}

\maketitle

\section{Introduction}
\label{sec:introduction}
The study of longitudinal vector boson scattering (VBS) processes remains a 
long-standing, yet unattained, milestone of high energy physics~\cite{Dicus:1990fz}.
The increase of the  scattering amplitude of longitudinal vector bosons with 
center-of-mass energy in absence of a Higgs boson, eventually violating 
unitarity~\cite{Veltman:1976rt,Lee:1977yc,Lee:1977eg}, led to the formulation
of the ``no-lose theorem'' for colliders of sufficiently high energies, postulating
that either a Higgs boson or some new physics beyond the Standard Model (SM)
be found~\cite{Gunion:1989we}. 
The discovery of a Higgs-like boson~\cite{Aad:2012tfa, Chatrchyan:2012ufa}
at the Large Hadron Collider (LHC) in 2012 heralded the first step toward
the study of the delicate interplay at work in longitudinal VBS. While the properties
of this Higgs boson thus far were found to be consistent with SM 
expectations~\cite{Khachatryan:2014kca, Khachatryan:2014jba,Aad:2015mxa,Aad:2015gba},
even small deviations in its vector boson couplings could give rise to an
increase of the scattering amplitude of longitudinal vector bosons with center-of-mass energy,
which in turn renders longitudinal VBS a sensitive probe for anomalous Higgs
couplings~\cite{Eboli:2006wa,Campbell:2015vwa}. Furthermore, a rich portfolio
of models beyond the SM predicts enhancements of VBS production through
extended Higgs sectors or other new resonances~\cite{Alboteanu:2008my,Godfrey:2010qb,
Espriu:2012ih,Chang:2013aya,Chiang:2014bia,Kilian:2014zja}. 

While experimentally VBS processes offer the distinct signature of a forward/backward pair of jets ($j$) which 
are well-separated in rapidity (``tagging'' jets) and exhibit a large invariant dijet mass, as well as the decay products of the
produced vector bosons, their measurement is challenging due to small cross sections.  Both ATLAS
and CMS have by now established VBS processes involving massive vector bosons ($V = W, Z$) in fully leptonic 
decay modes, successfully separating the desired purely electroweak production from strong (QCD-induced) $VVjj$ 
production and other background processes~\cite{Aaboud:2018ddq,Sirunyan:2020gyx,Aaboud:2019nmv,Sirunyan:2017ret,
Aad:2020zbq,Sirunyan:2020alo}. Studies of semileptonic VBS final states where one vector boson decays hadronically -- albeit 
benefitting from the large $V$ hadronic branching fraction compared to the leptonic decays -- thus far were unable to firmly 
establish the SM signal due to increased background levels, but have proven to provide excellent sensitivity to anomalous 
couplings~\cite{Aaboud:2016uuk,Sirunyan:2019der,Aad:2019xxo}. Observation of the semileptonic $W V jj$ VBS process is expected to 
be achievable at the LHC with an integrated luminosity of 300~fb$^{-1}$ at $\sqrt{s} = 14$~TeV~\cite{ATLAS:2018ocj}.

Measuring {\em longitudinal} VBS processes is further complicated by the difficulty of separating longitudinal states from transverse ones.
First studies of $VV$ and $VVjj$ LHC data in fully leptonic vector boson decay modes explore the possibilities of 
extracting cross sections for longitudinally polarized vector bosons, but their sensitivity is still insufficient in the currently 
available datasets to access longitudinal VBS~\cite{Aaboud:2019gxl,Sirunyan:2020gvn}. 
Projections to the High-Luminosity LHC (HL-LHC), providing proton--proton collisions at a center-of-mass
energy of $\sqrt{s} = 14$~TeV with an integrated luminosity of 3000~fb$^{-1}$, show $W^\pm W^\pm jj$ as the most promising
channel to establish longitudinal VBS in fully leptonic vector boson decay modes, but none of the $W^\pm W^\pm jj$, $W Zjj$ 
and $ZZ jj$ processes studied is predicted to reach a significance of 3 standard deviations at a single experiment~\cite{CMSCollaboration:2015zni,
CMS:2018mbt,CMS:2018ylh,ATLAS:2018tav,CMS:2018zxa,ATLAS:2018uld}.
Using more sophisticated analysis techniques such as  deep machine learning, the sensitivity for longitudinal VBS can be significantly increased,
as demonstrated for $W^\pm W^\pm jj$ in~\cite{Searcy:2015apa} for the leptonic decay channel.

The sensitivity to longitudinal VBS in {\em semileptonic} final states has been explored much less. While some studies
exist for $\sqrt{s} = 13$~TeV~\cite{Grossi:2020orx} or $\sqrt{s} = 27$~TeV~\cite{Cavaliere:2018zcf},
the sensitivity to longitudinal VBS in semileptonic final states at the HL-LHC has not been assessed -- a gap that this
paper is addressing. We focus on the $W V jj$ channel where one $W$ boson decays into a charged lepton (an electron or muon, denoted 
by $\ell$) and an (anti-) neutrino $\nu$, while the other massive vector boson $V$ is considered to decay into a pair of quarks which we require 
to be reconstructed as a merged, large-radius jet ($J$), leading to an $\ell\nu J jj$ final state. To enable full reconstruction of the event kinematics, 
the neutrino four-vector is recreated by imposing a $W$ boson mass constraint in the lepton neutrino system. The expected significant impact
of pileup at the HL-LHC is mitigated by using track-based observables, and jet substructure techniques are deployed to improve $V$ boson reconstruction.
As both the resolved $W V jj$ channel (where the hadronic $V$ decay is reconstructed via two separate small-radius jets) and $ZV jj$ semileptonic 
final states will contribute to establishing longitudinal VBS in semileptonic final states, our results can be seen as a lower limit on the expected sensitivity
at the HL-LHC.

\section{Simulation Samples}
\label{sec:samples}
Electroweak $WVjj$ production includes contributions from the $W^\pm W^\mp jj$, 
$W^\pm W^\pm jj$ and $W Z jj$ VBS processes, which are modeled with 
\MGMCatNLO~2.7.3~\cite{Alwall:2014hca},
interfaced to \PYTHIA~8.243~\cite{Sjostrand:2007gs} for parton showering and hadronization. 
These samples are generated with two on-shell vector bosons, with one $W$ boson decaying leptonically ($W \rightarrow \ell\nu$), and the other massive vector boson decaying hadronically.
The contribution from triboson processes is also included,
but negligible in the phase space studied (see Sec.~\ref{sec:evSel}).
Four different polarization states are produced at leading order in QCD: both bosons are longitudinally polarized ($W_{L}V_{L}$), both transversely polarized ($W_{T}V_{T}$),
or a mixture ($W_{T}V_{L}$ and $W_{L}V_{T}$). 
These polarized samples are simulated with the helicity eigenstates defined in the $WV$ center-of-mass reference frame~\cite{BuarqueFranzosi:2019boy}.
For this analysis focused on $W_{L}V_{L}$ production, the signal is referred to as VBS $W_{L}V_{L}$, while the other polarization states ($W_{T}V_{T}$, $W_{T}V_{L}$, 
and $W_{L}V_{T}$) are referred to as the VBS $WV$ background.

The main background contributions for this analysis are the production of a $W$ boson in association with jets and top-quark pair production. 
The $W$+jets samples are simulated using CKKW-L merging~\cite{Catani:2001cc,Lonnblad:2001iq} with up to four partons at leading order in QCD using \MadGraph~2.7.3.
The top-quark pair production sample is generated using \MadGraph~2.7.3 at next-to-leading order in QCD, and the top quarks are decayed using \MadSpin~\cite{Artoisenet:2012st} in order to preserve the spin correlations for top-quark production and decay.
\Pythia~8.243 is used for parton showering and hadronization for all background samples.
The contribution of QCD-induced $VW$jj processes in our signal region is found to be a factor of 100 smaller than the $W$+jets background, and a factor of two smaller than the VBS $VW$ EW-induced backgrounds, and is not considered further in our studies.

A parton level event filter of $H_T > 200$ GeV is used to enhance the statistical power of the Monte Carlo (MC) samples in the phase space studied, where $H_T$ is the sum of the transverse momentum ($p_T$) of all partons.
Leptons and partons are also required to satisfy $p_T > 10$ GeV at the generator level.
This filter is found to be fully efficient for the concerned phase space in this study (see Sec.~\ref{sec:evSel}).
Table~\ref{tab:mc} summarizes the simulated MC samples.
The number of events generated in particular for the background processes is driven by the requirement that there be 
no empty bins in the discriminant used to determine the analysis sensitivity (see Sec.~\ref{sec:sensitivity}), hence 
avoiding any extrapolations across empty bins.
All signal and background processes are reconstructed using a generic detector in the Delphes simulation framework~\cite{deFavereau:2013fsa},
modeled after the ATLAS detector in the HL-LHC~\cite{Azzi:2019yne}.

\begin{table}[ht]
\centering
\caption{Overview of the simulated MC samples.}
\label{tab:mc}
\begin{tabular}[t]{lccc}
\hline
Process & Accuracy & Cross section [fb] & Number of events \\
\hline
\WpWmVBS & LO & 0.325 & 2.7e6 \\
\WZVBS & LO & 0.114 & 3.9e5 \\
\ssWWVBS & LO & 0.114 & 5.89e5 \\
\hline
$W$+jets & LO & 1185 & 1.09e7 \\
\ttbar & NLO & 374 & 8.64e6 \\
\hline
\end{tabular}
\end{table} 
\section{Event selection}
\label{sec:evSel}
Events from VBS $WV$ production exhibit several distinct characteristics which may be used in the event selection.
In the semileptonic decay, the event contains one lepton and missing transverse momentum $E_{\text{T}}^{\text{miss}}$ from the leptonic $W$ boson decay, and either two jets or a large-radius jet from the hadronic $V$ decay. In addition to the $V$ boson decay products, there are two forward jets from the VBS production, which are referred to as the ``tagging'' jets.

As detailed in Table~\ref{tab:evsel}, a loose selection is applied to the events, based on the expected reconstruction capabilities at the HL-LHC~\cite{Atlas:2019qfx}. 
In order to select the leptonically decaying $W$ boson, each event is required to have an electron or muon with $p_T > 20$ GeV and pseudorapidity $|\eta_{\ell}|<$4.0, and to contain no other leptons with $p_{T} > 7$ GeV. 

Our study focuses on the case where the hadronically decaying $V$ boson candidate can be reconstructed as a single large-radius jet $J$.
The inclusion of the resolved case where the decay products are reconstructed as separate jets would improve the significance of these results, but is not considered due to the combinatoric challenges in assigning tagging- and $V$-decay jets.
Jets are clustered with $\textsc{FastJet}$~\cite{Cacciari:2011ma} using the anti-$k_t$ algorithm~\cite{Cacciari:2008gp} with radius parameter of $R=1.0$, using ``particle flow objects'' as inputs to the jet reconstruction algorithm. These particle flow inputs combine information from the tracker and calorimeter in order to provide better resolution for object reconstruction.
The large-radius jets are groomed using the soft-drop grooming algorithm, with $\beta=1.0$, and $z_{cut}=0.1$~\cite{Larkoski:2014wba} in order to reduce effects due to multiple simultaneous $pp$ collisions (pileup) and the underlying event, and to improve sensitivity of the $V$ boson reconstruction.
The large-radius jet is required to have $p_{T} > 200$ GeV in order to reconstruct both decay products within a single jet, and $|\eta| < $  4.0, and it is required to be isolated from the lepton by $\Delta R_{\ell, J} > 1.0$.
If multiple large-radius jets are reconstructed, the highest-$p_{T}$ jet is selected. 
After the jet is selected, its mass is required to satisfy $40 < m_{J} < 180$ GeV.

Missing transverse momentum is reconstructed as the negative sum of the transverse momentum of all particle-flow objects within $|\eta| < 5.0$, 
and is required to be greater than 80 GeV to reduce QCD background contributions.

The two quarks produced in the VBS production are reconstructed using small-radius ($R=0.4$) jets.
These jets are required to have $p_{T} > 30$ GeV and $|\eta| < 4.0$, and they must be isolated from the selected large-radius jet by $\Delta R > 1.4$, and from the lepton by $\Delta R > 0.4$.
The two jets that maximize the dijet invariant mass ($m_{j_{1}j_{2}}$) and are in opposite hemispheres ($\eta_{j_{1}} \cdot \eta_{j_{2}} < 0$) are identified as the tagging jets,
and events are required to have $m_{j_{1}j_{2}} > 800$ GeV to reduce the background contributions. 
To lower contributions from top-quark pair production, the event is required to have no b-tagged jets outside of the selected large-R jet.

\begin{table}[ht]
\centering
\caption{The event selection.}
\label{tab:evsel}
\begin{tabular}[t]{ll}
\hline
Object & Selection\\
\hline
Lepton                       & $p_{T,\ell}>$20 GeV \\
                             & $|\eta_{\ell}|<$4.0 \\
                             & No other leptons with $p_{T,\ell}>$7 GeV  \\
Large-R jet $J$                 & $p_{T,J}>$200 GeV\\
                             & $|\eta_{J}|<$4.0 \\
                             & 40 GeV $< m_{J}<$180 GeV\\
$E_{\text{T}}^{\text{miss}}$ & $E_{\text{T}}^{\text{miss}} >$ 80 GeV\\
tagging jets ($j_{1}$, $j_{2}$)    & $p_{T,j} >$ 30 GeV \\
                             & $|\eta_{j}| <$ 4.0  \\
                             & $\Delta R_{j, J} >$ 1.4, $\Delta R_{j, \ell} >$ 0.4  \\
                             & $m_{j_{1}j_{2}} > $ 800 GeV \\
                             & $\eta_{j_{1}} \cdot \eta_{j_{2}} < $ 0 \\
                             & No b-tagged jets in the event with $\Delta R_{J, j} >$ 1.0 \\
\hline
\end{tabular}
\end{table}

\subsection{$V$ boson reconstruction}
After above event selection, both of the $V$ bosons are reconstructed at detector level.
The large-radius jet serves as a proxy for the hadronically decaying $V$ boson. 
Since the jet has been groomed with the soft-drop algorithm, the two associated subjets which pass the soft-drop condition are natural proxies for the decay products of the $V$ boson.

The leptonically decaying $W$ boson is fully reconstructed using the lepton and $E_{\text{T}}^{\text{miss}}$, using the $W$ boson mass to fully constrain the kinematics,
with the assumption that the $E_{\text{T}}^{\text{miss}}$ arises solely from the neutrino.
The neutrino transverse momentum is taken to be the $E_{\text{T}}^{\text{miss}}$, and the longitudinal component is solved for by assuming the $W$ boson is on-shell, and that the charged lepton is massless. The result of this is a second order polynomial with two solutions. In cases where there are no real solutions, the longitudinal momentum is taken to be the real component of the solution.
In cases where there are two real solutions, the solution with the smaller longitudinal momentum is taken, which produces the correct result in around 65\% of generated events.

\subsection{Polarization}

In the $V$ boson rest frame, the decay products of the $V$ boson will be back-to-back, and can be characterized based on the angle $\theta^{*}$ between the $V$ boson direction and the decay product direction. The $V$-boson differential cross section depends on the polarization fractions as

\begin{equation}
\frac{d\sigma}{d \mathrm{cos} \theta^{*}} \propto
 \frac{3}{8} f_{-}(1 \mp \mathrm{cos} \theta^{*})^2 +
\frac{3}{8} f_{+}(1 \pm \mathrm{cos} \theta^{*})^2 + 
\frac{3}{4} f_{L} (1-\mathrm{cos}^2 \theta^{*}), \mathrm{for W^{\pm}},
\end{equation}

where $f_{-}$, $f_{+}$, and $f_{L}$ are the fractions of events where the $V$ boson polarization is $-1$, $+1$, and 0, respectively.
Similarly, in the laboratory frame, the decay products for the longitudinally polarized $V$ bosons will tend to be more balanced in $p_T$, and less balanced for transversely polarized $V$ bosons.
Consequently, the momentum balance of the leptonic decay products, or  $z_{g, \ell} = p_{T,\ell} / p_{T, W}$, and $cos(\theta^{*})_\ell$ are sensitive variables to the $W$ boson polarization.

Similar variables may be defined for the hadronic case as well, using the large-R jet and its two subjets as proxies for the hadronically decaying $V$ boson and its decay products. 

\begin{figure}[thb]
  \centering
  \includegraphics[width=0.48\textwidth]{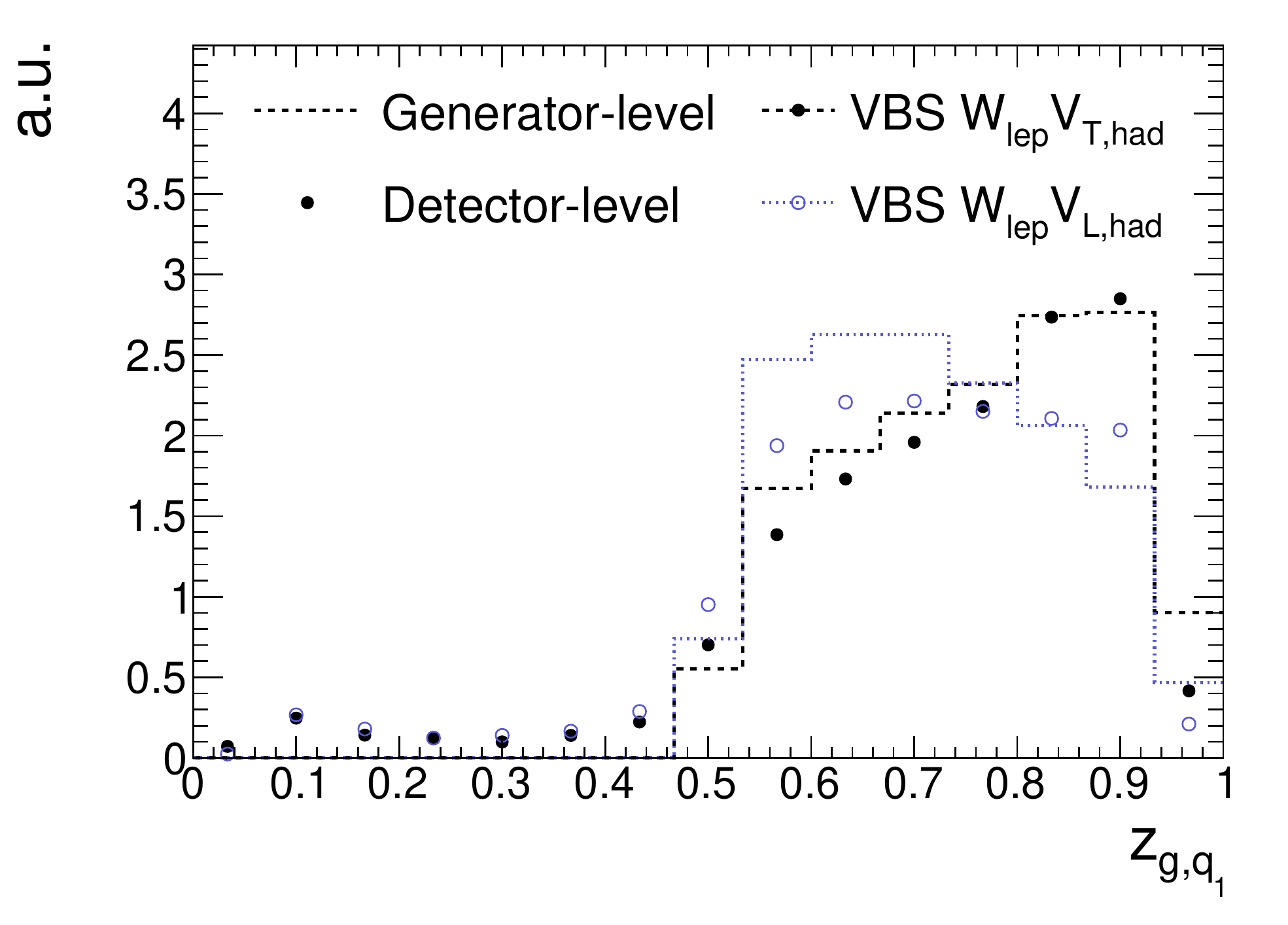}
  \includegraphics[width=0.48\textwidth]{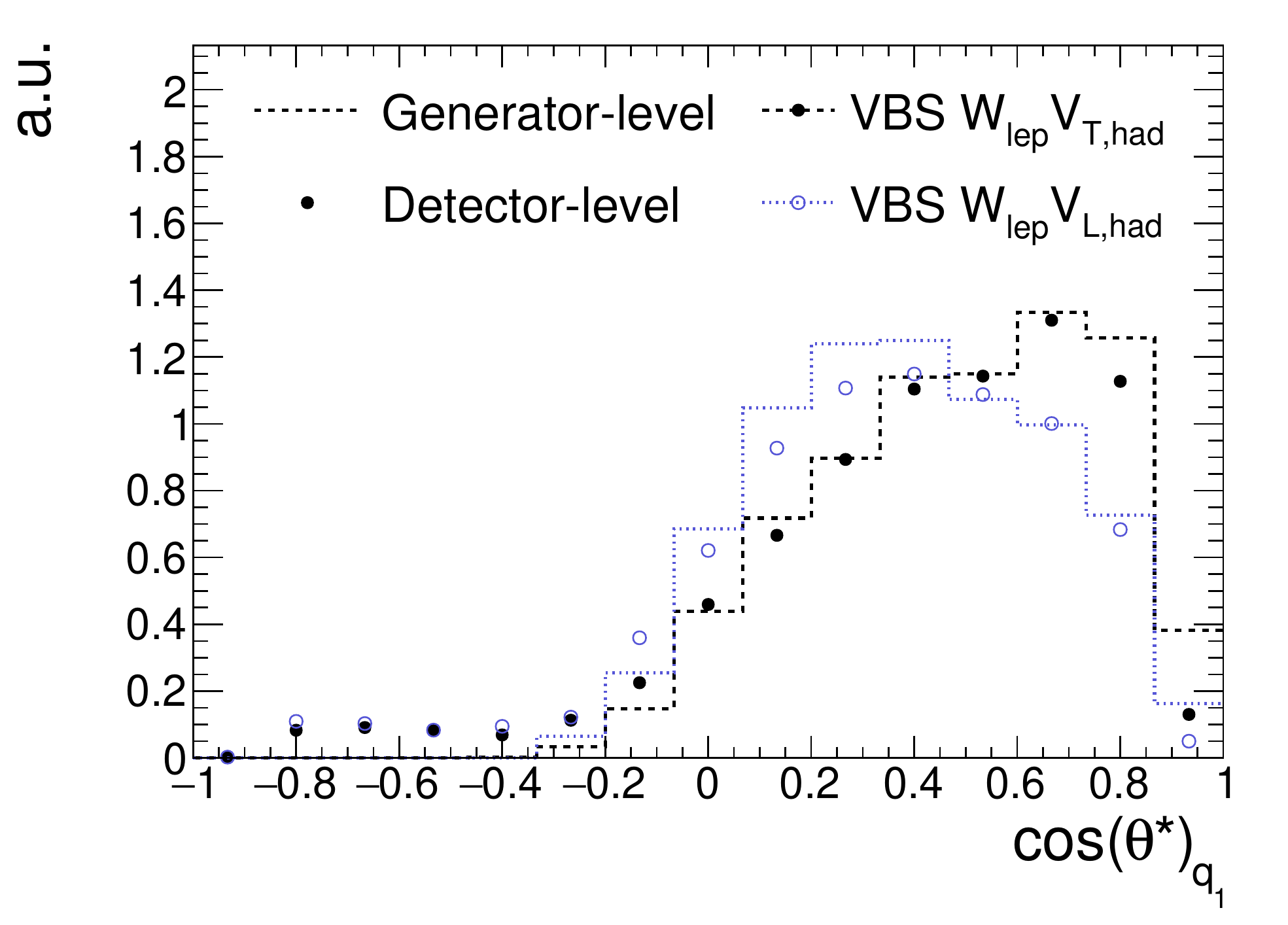}
  \includegraphics[width=0.48\textwidth]{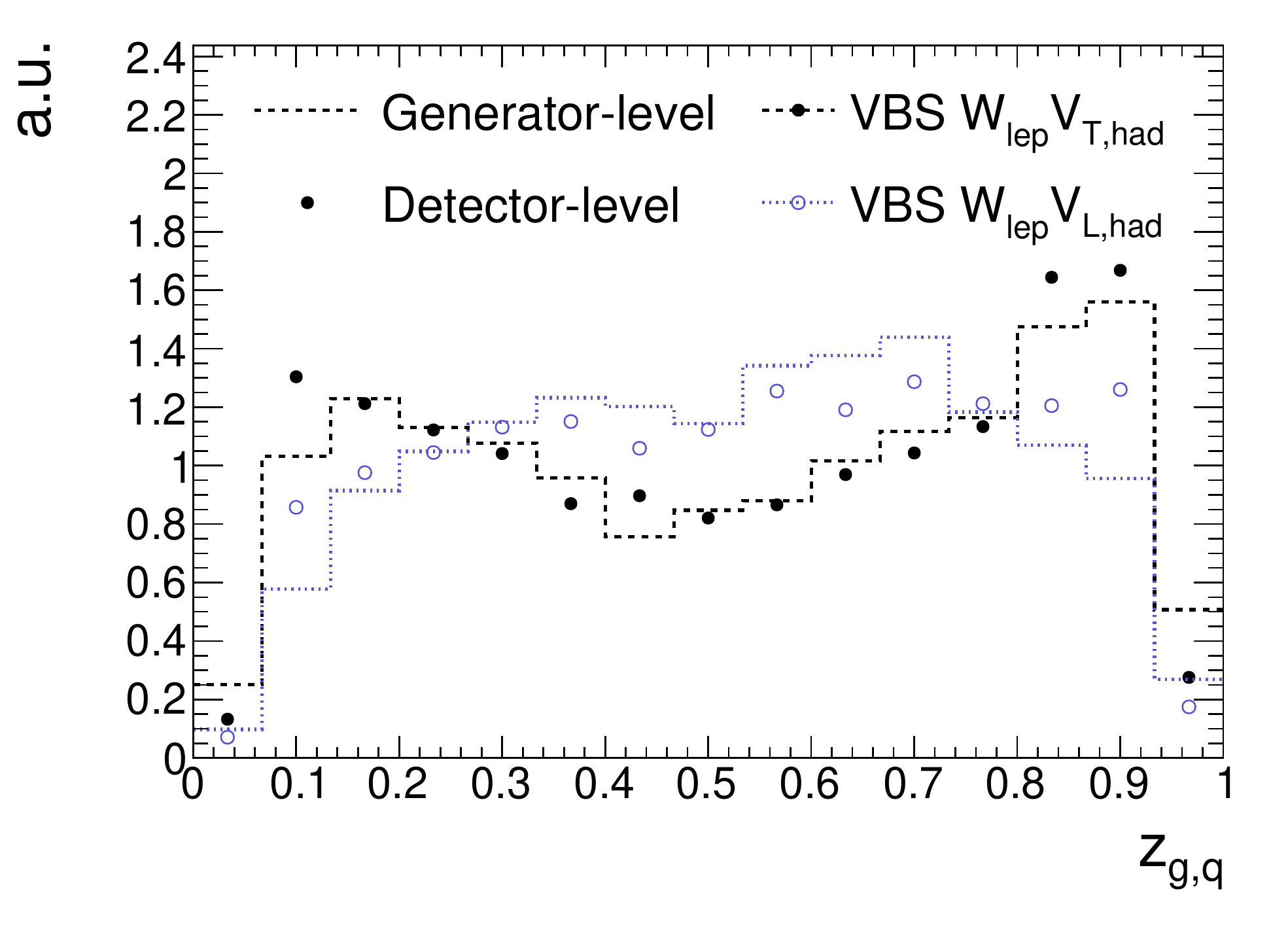}
  \includegraphics[width=0.48\textwidth]{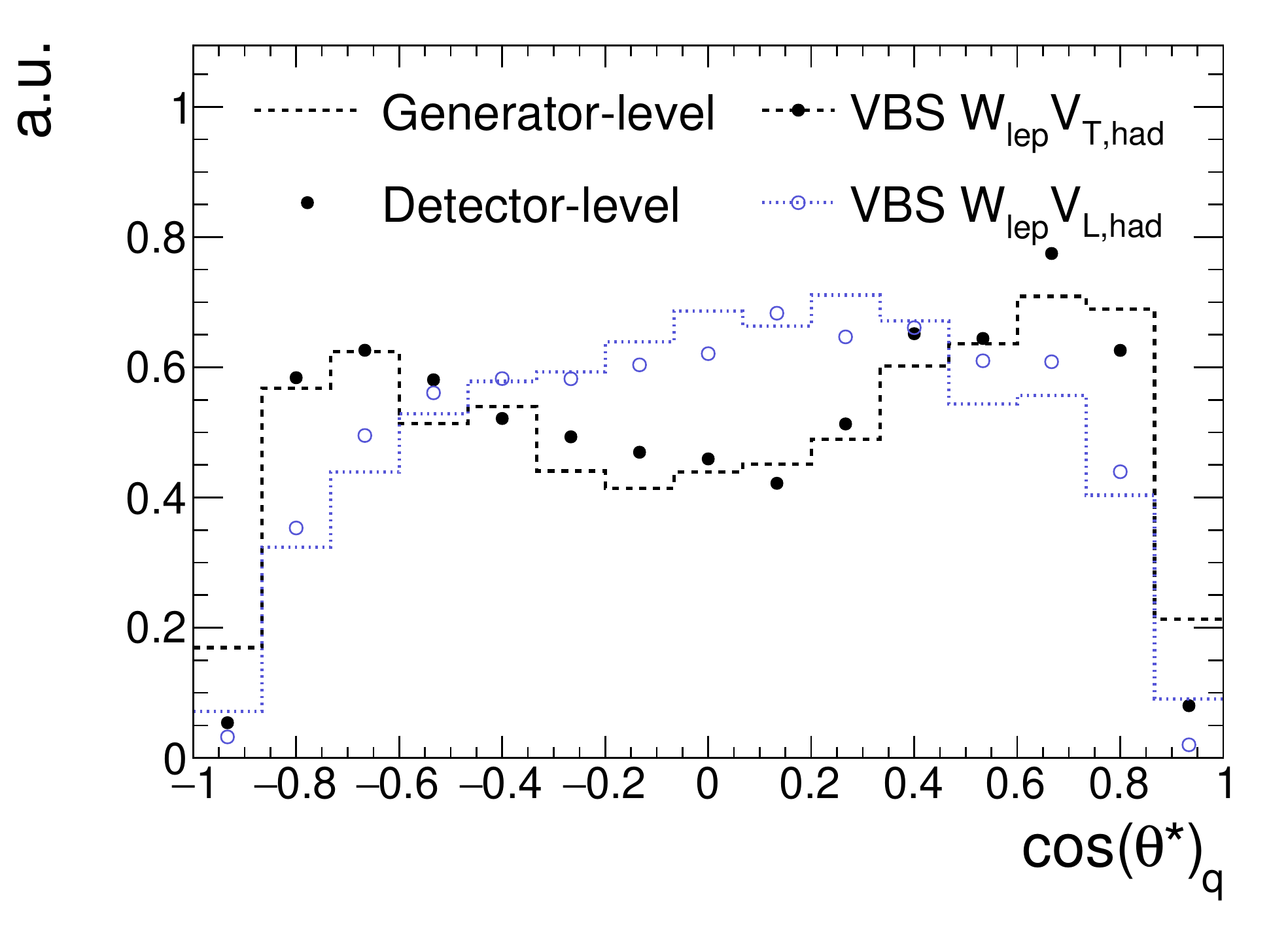}
  \caption{Comparison of (left) the $p_T$ balance of the hadronically decaying $V$ decay products, and (right) the $cos(\theta^{*})$ distribution for generator and detector level reconstruction for longitudinally and transversely polarized $V$ bosons, using (top) the leading $p_T$ subjet $q_1$, and (bottom) the quark $q$ from the $V$ boson decay.}
  \label{fig:wbos}
\end{figure}

Using MC generator truth information, the decay products are distinguishable as quark ($q$) and anti-quark ($\bar{q}$), and we can use for example $cos(\theta^{*})_{q}$ (defined using the angle between the $V$ boson direction and the quark $q$ from the $V$ boson decay), and $z_{g, q} = p_{T, q} / p_{T, V}$ as polarization-sensitive observables without introducing a kinematical bias, albeit not reconstructable in data. At detector level, the two subjets are only distinguishable by their kinematics, and denoting the leading $p_T$ subjet $q_1$ and the subleading $p_T$ subjet $q_2$, we can define e.g. $cos(\theta^{*})_{q_{1}}$ and $z_{g, q_{1}} = p_{T,q_{1}} / p_{T, V}$ accordingly. While this biases the kinematics (as illustrated in Fig.~\ref{fig:wbos}), these observables are accessible with the detector. 

To validate the assumption that the subjets are good proxies for the $V$ boson decay products, above observables -- both using $q$ and using $q_1$ -- are studied with a few additional requirements.
To reduce the contributions of events where the $V$ boson decay products are not contained within a single large-radius jet, the $V$ boson is required to be matched to the selected large-radius jet with $\Delta R_{V, J} < 0.4$.
The subjets are ordered with the same $\eta$-ordering as the generator-level decay products, to avoid any bias from a direct matching of the subjets and the generator-level decay products.
A comparison of the generator-level and detector-level distributions for cos$\theta^*$ and $z_g$ is shown for the hadronically decaying $V$ boson in Fig.~\ref{fig:wbos}, which demonstrates that the subjets indeed are good proxies for the $V$ boson decay products, and can be used to distinguish between the different polarization states of the $V$ boson.

\begin{figure}[thb]
  \centering
  \includegraphics[width=0.48\textwidth]{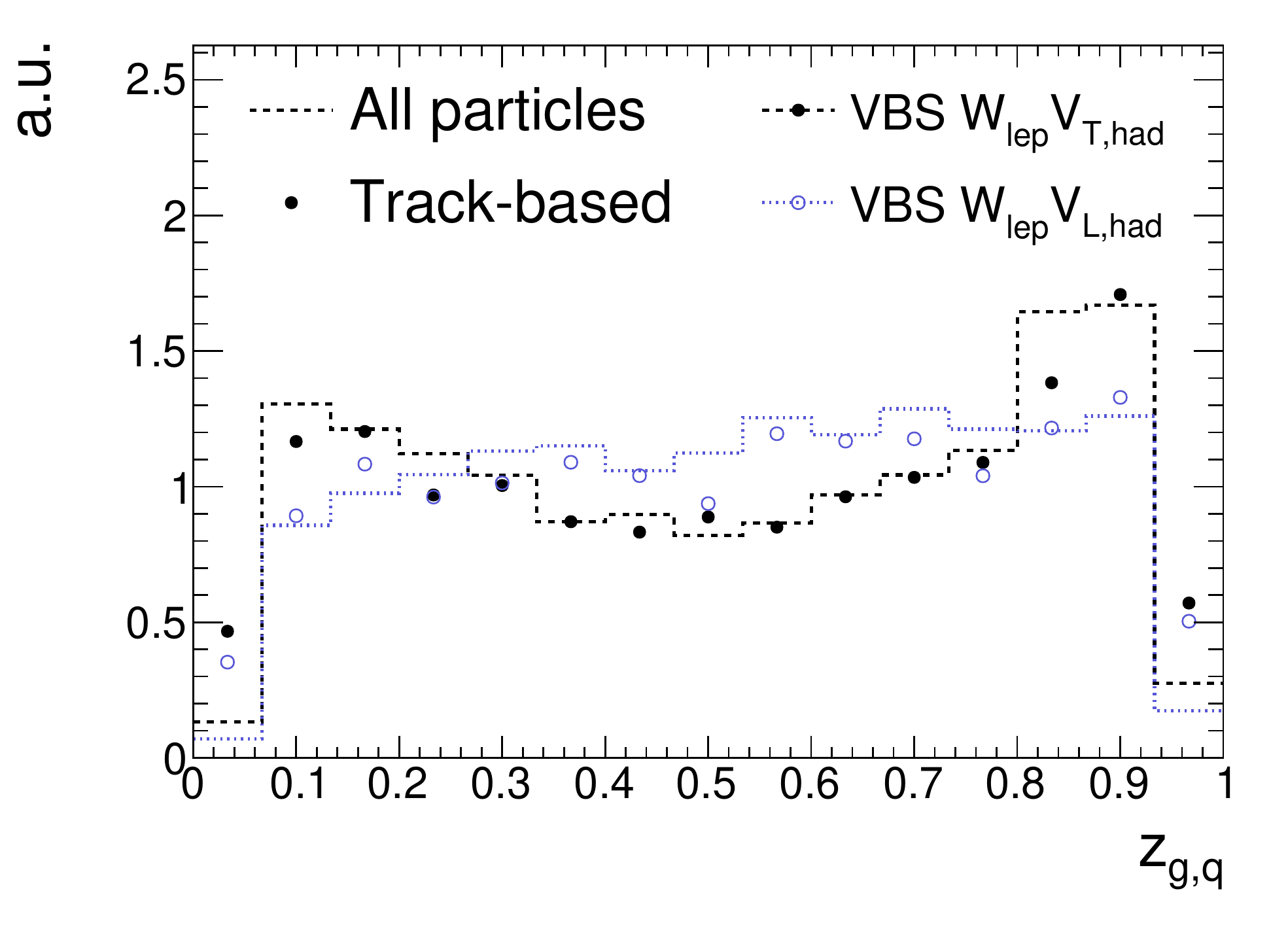} 
  \includegraphics[width=0.48\textwidth]{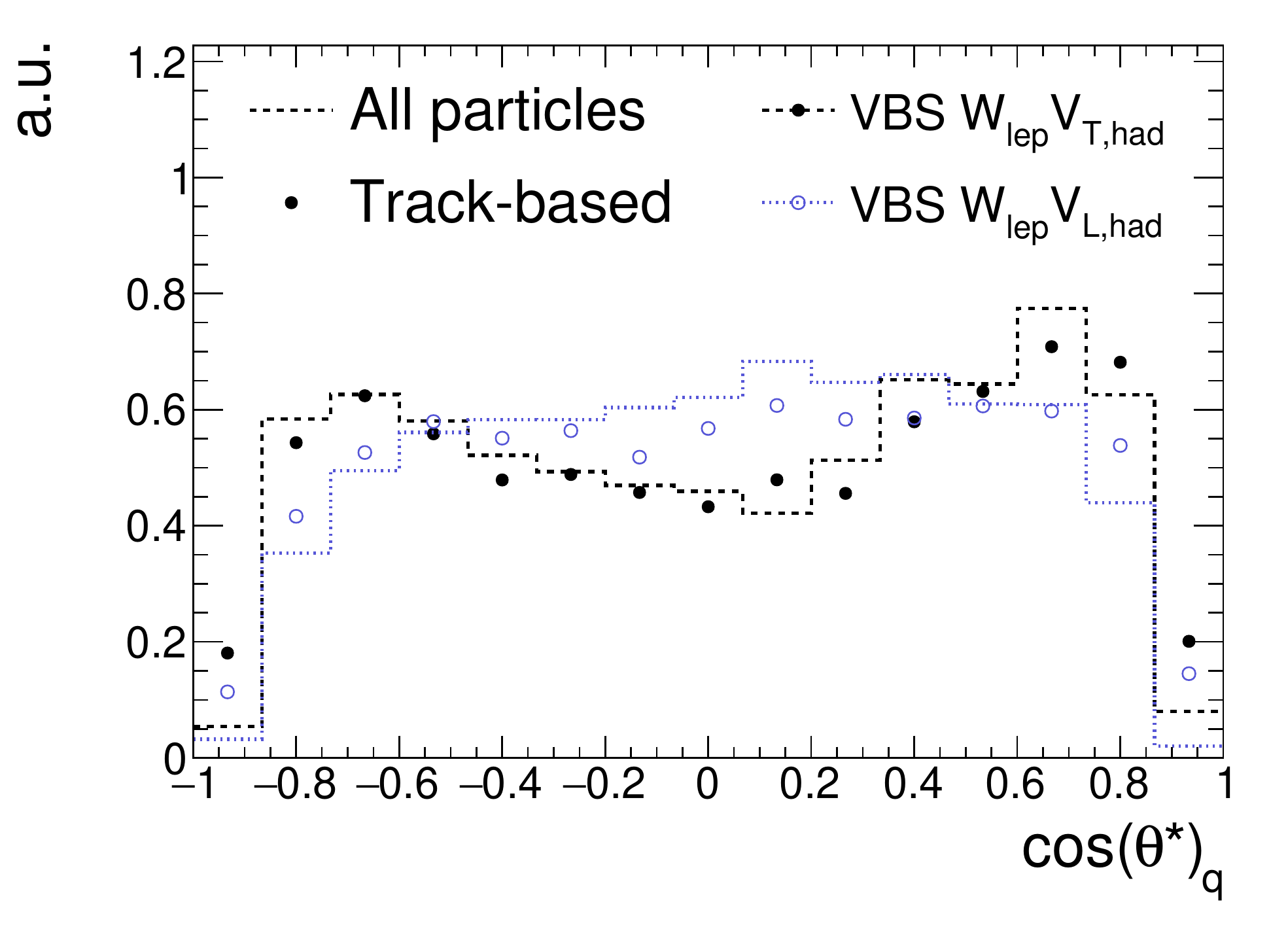} 
  \caption{Comparison of (left) The $p_T$ balance of the hadronically decaying $V$ decay products, and (right) the $cos(\theta^{*})$ distribution for particle-flow-based and track-based substructure reconstruction for longitudinally and transversely polarized $V$ bosons.}
  \label{fig:trackVall}
\end{figure}

At the HL-LHC, reconstruction is complicated by the impact of radiation from pileup on these observables.
In particular, jet substructure is sensitive to the wide-angle, low-$p_T$ particles associated with pileup.
Particle flow objects include calorimeter measurements, where pileup is difficult to separate from the hard-scatter collision, while for tracks, pileup may be removed based on the primary vertex association. In order to mitigate their pileup sensitivity, jet substructure observables can be calculated using tracks as inputs rather than using particle-flow objects. 
To reconstruct these track-based observables, tracks are associated to a large-R jet using a $\Delta R < 1.0$ matching. These tracks are then clustered and groomed using the same algorithms as the particle-flow jets. Consequently, each substructure observable may be calculated using either the particle-flow constituents of the jet, or the groomed tracks associated to the jet.
As illustrated in Fig.~\ref{fig:trackVall}, track-based observables are able to capture similar information as the particle-flow observables and hence are used in our analysis from here on for substructure observables whose pileup sensitivity has not been studied in detail, namely any substructure observables which are not the jet mass or ratios of energy correlation functions such as $d_2$~\cite{Larkoski:2015kga}.

\begin{figure}[thb]
  \centering
  \includegraphics[width=0.48\textwidth]{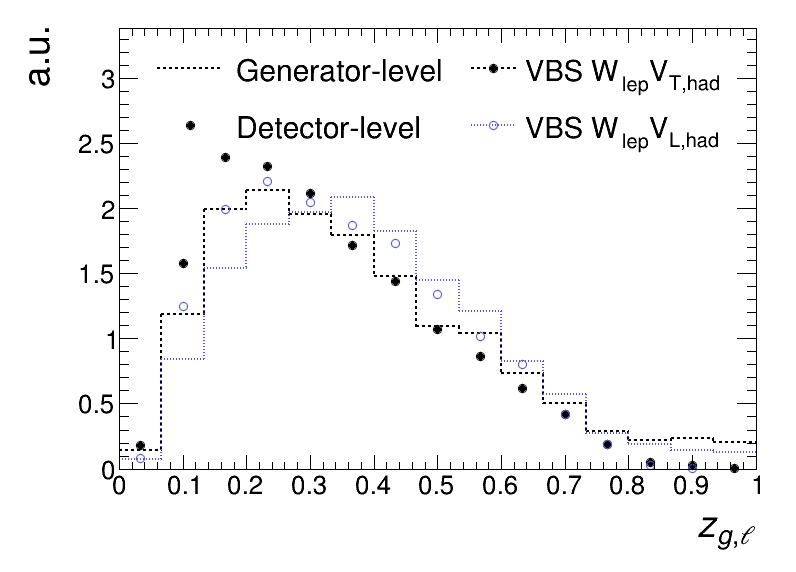}
  \includegraphics[width=0.48\textwidth]{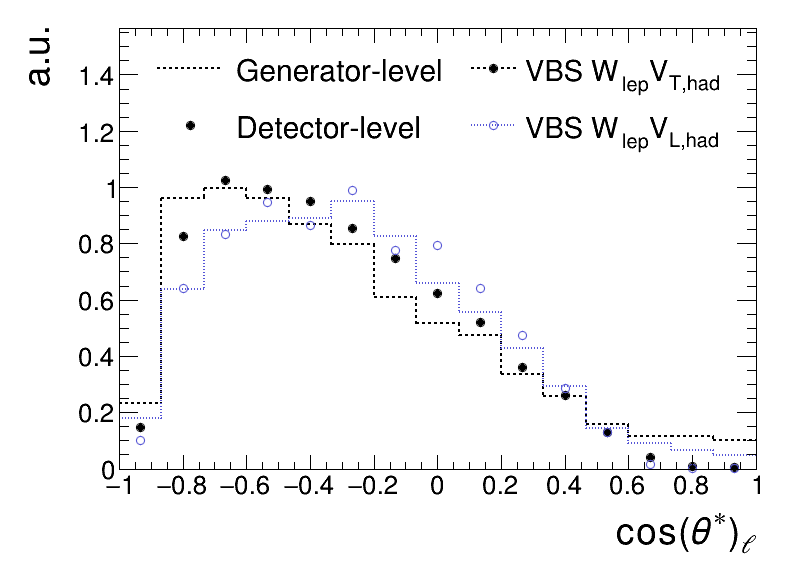}
  \caption{Comparison of (left) the $p_T$ balance of the leptonically decaying $W$ decay products, and (right) the $cos(\theta^{*})$ distribution for generator and detector level reconstruction for longitudinally and transversely polarized $W$ bosons.}
  \label{fig:wlep}
\end{figure}

Since the leptonically decaying $W$ boson is fully reconstructed, it is also possible to define similar observables using the lepton and the reconstructed $W$ boson. 
The corresponding results are shown for events with particle-level $E_{\text{T}}^{\text{miss}}>80$ GeV in Fig.~\ref{fig:wlep}, illustrating that the reconstructed $W$ boson decay behaves similarly to the generator-level $W$ boson. 
In the transversely polarized case, the lepton tends to have a $p_{T}$ smaller than the neutrino. This is a result of the $E_{\text{T}}^{\text{miss}}$ cut, 
which biases the relative momenta of the $W$ decay products. 

\section{Signal Extraction}
\label{sec:tagger}
\begin{figure}[thb]
  \centering
  \includegraphics[width=0.48\textwidth]{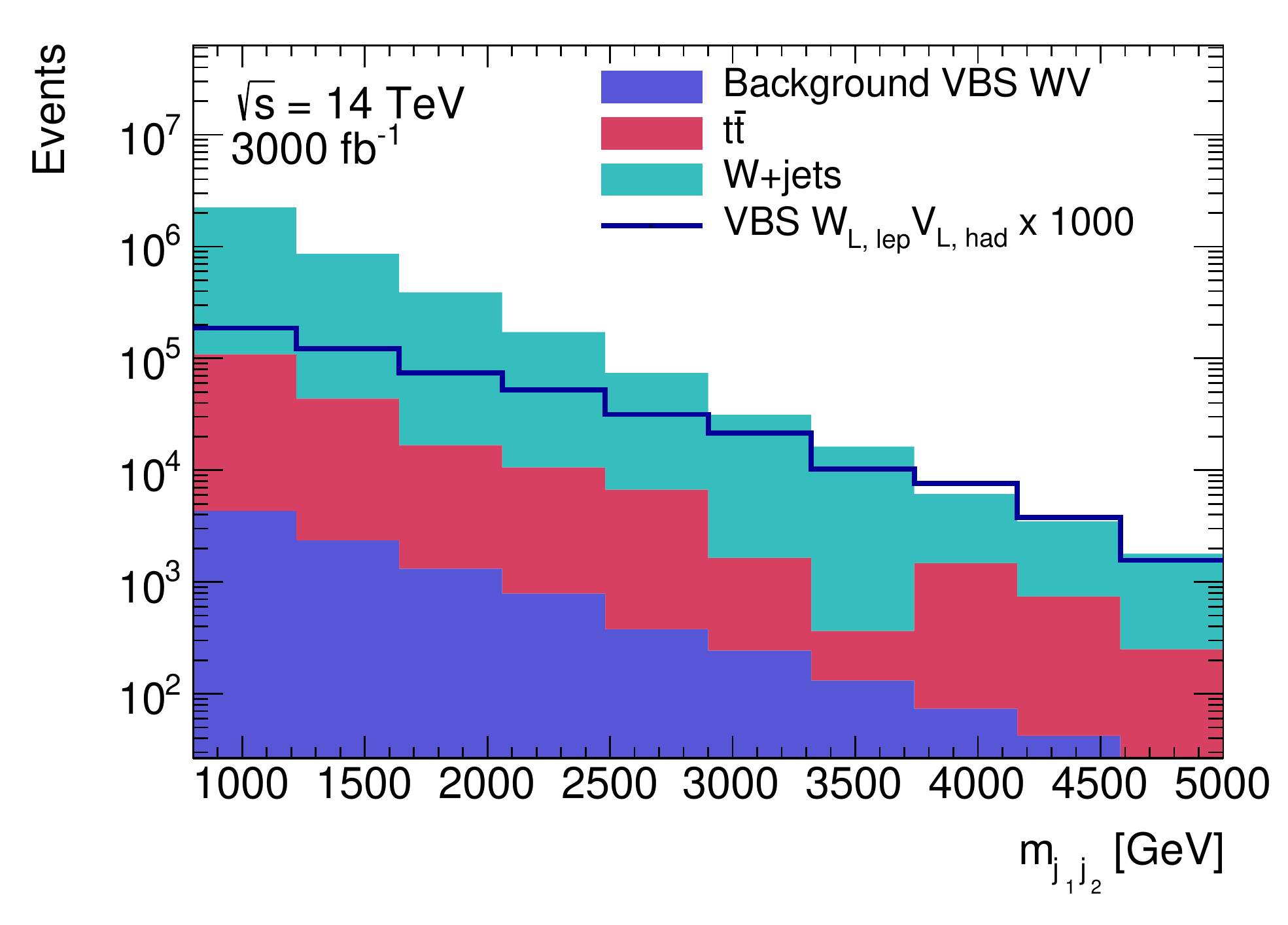}
  \includegraphics[width=0.48\textwidth]{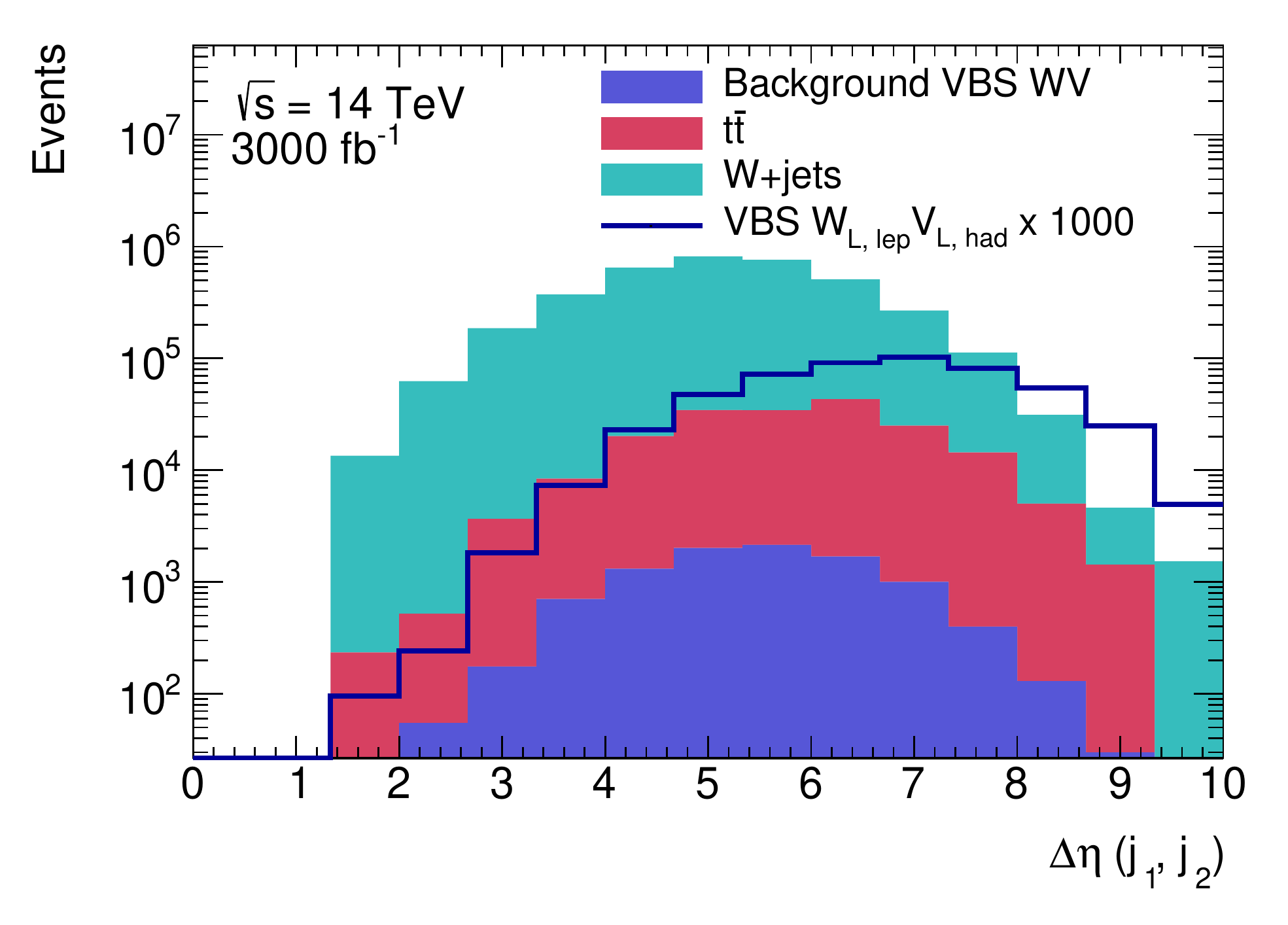}
  \includegraphics[width=0.48\textwidth]{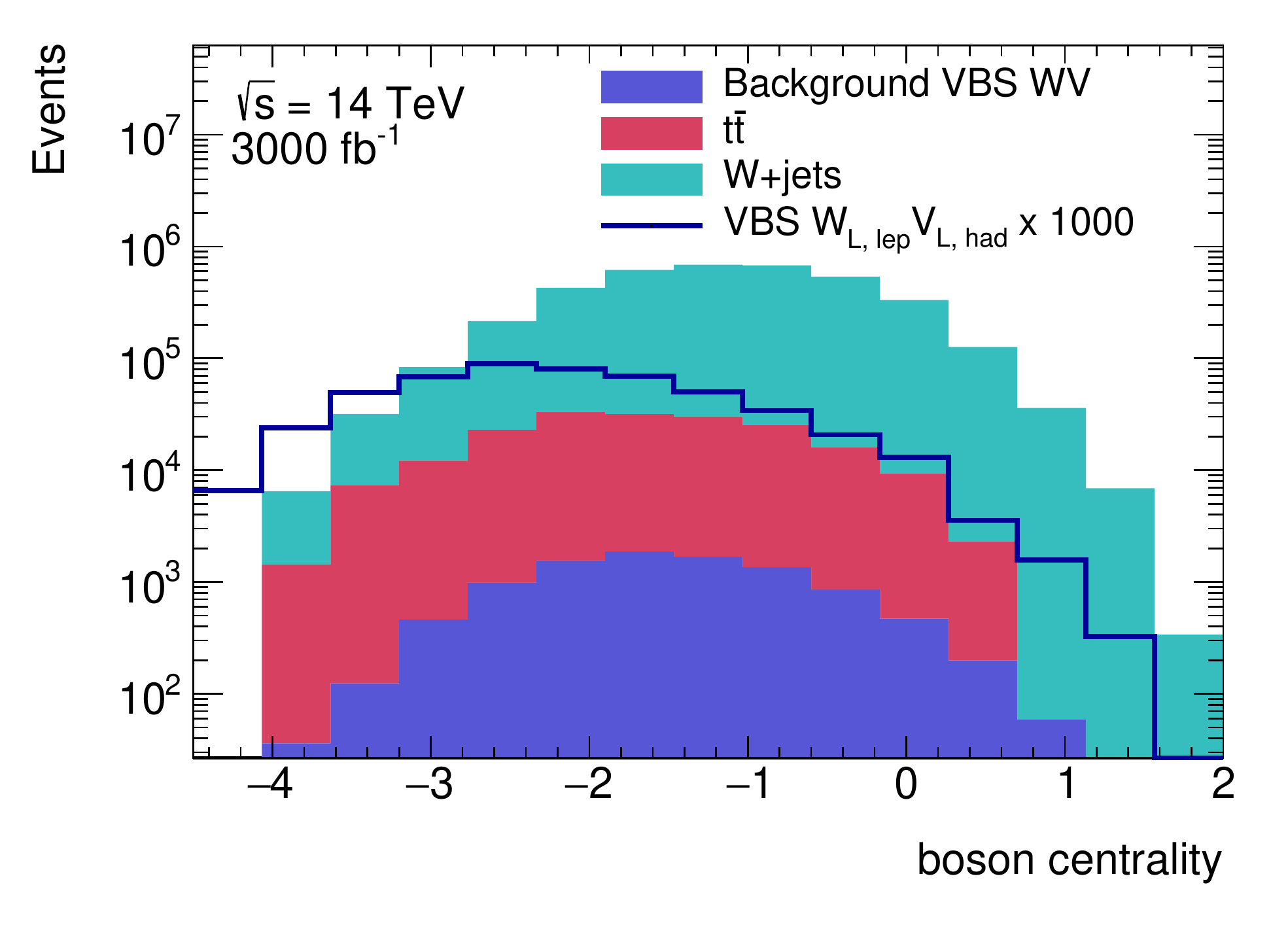}
  \includegraphics[width=0.48\textwidth]{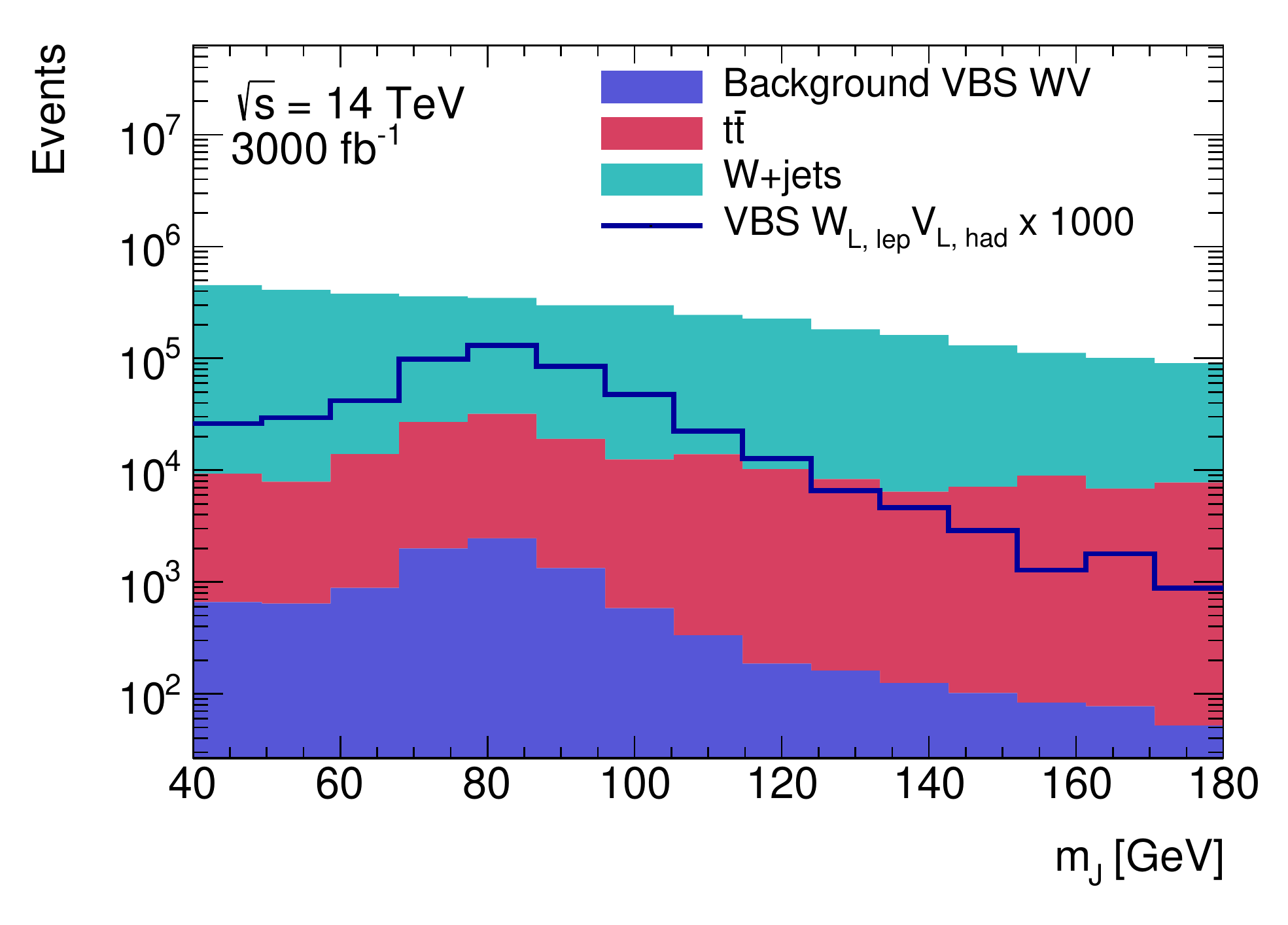}
  \caption{Kinematic distributions of the $W_{L}V_{L}$ signal compared to the three main background processes after the event selection for an integrated luminosity of 3000 fb$^{-1}$.}
  \label{fig:eventyields}
\end{figure}

Three main background processes need to be considered to extract the longitudinal VBS signal: $W$+jets and top-quark pair production, as well as the VBS non-$W_{L}V_{L}$ polarization states. To illustrate the initial signal-to-background ratio, the event yields of signal and background for 3000~fb$^{-1}$ of data after applying the event selection are shown in Fig.~\ref{fig:eventyields} for several observables.
No single observable offers sufficient background reduction on its own, but by combining multiple observables in a neural network, the background reduction can be significantly improved.

Each different background has unique characteristics which may be used to distinguish it from the $W_{L}V_{L}$ signal process: 
\begin{itemize}
  \item The background VBS $WV$ events have a similar topology, but differ for variables sensitive to the polarization states. 
  \item The $W$+jets background does not contain a hadronically decaying $V$ boson, and the tagging jets will tend to be more central.
  \item The top-quark pair production background contains a hadronically decaying $W$ boson, and will tend to have more (heavy flavor) jets in the event. 
\end{itemize}
Because of this, it is difficult to train a tagger to effectively distinguish between the $W_{L}V_{L}$ events and all background processes.
In order to improve analysis sensitivity, a multiclass tagger is trained to identify four different classes of events: 
the signal (VBS $W_{L}V_{L}$), the other (background) polarization states of VBS $WV$, $W$+jets, and top-quark pair production.

\begin{figure}[thb]
  \centering
  \includegraphics[width=0.31\textwidth]{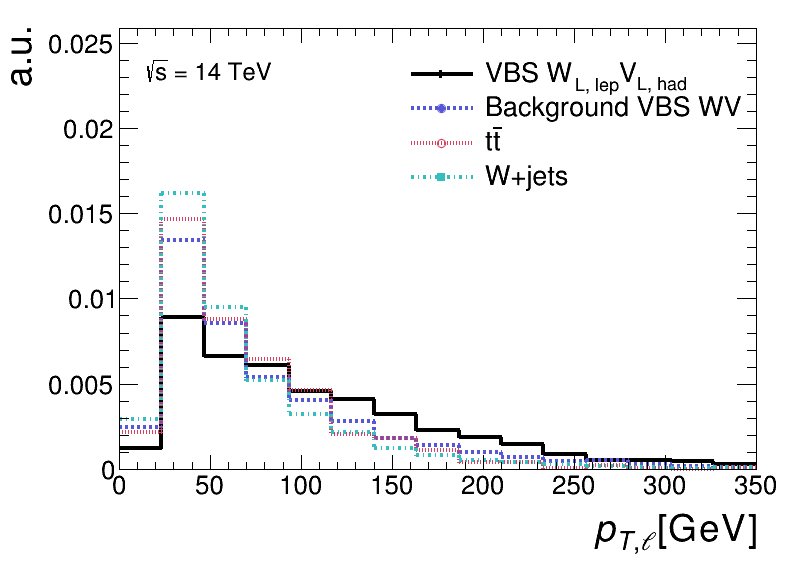}  
  \includegraphics[width=0.31\textwidth]{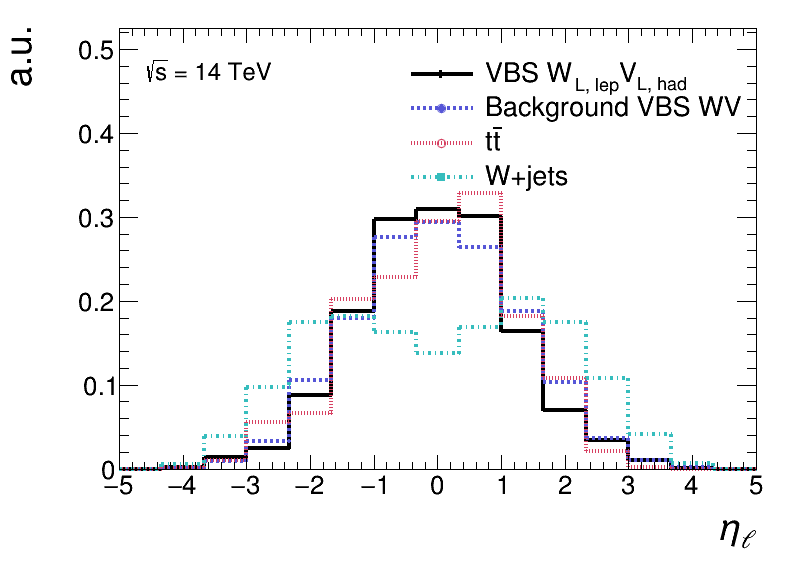}
  \includegraphics[width=0.31\textwidth]{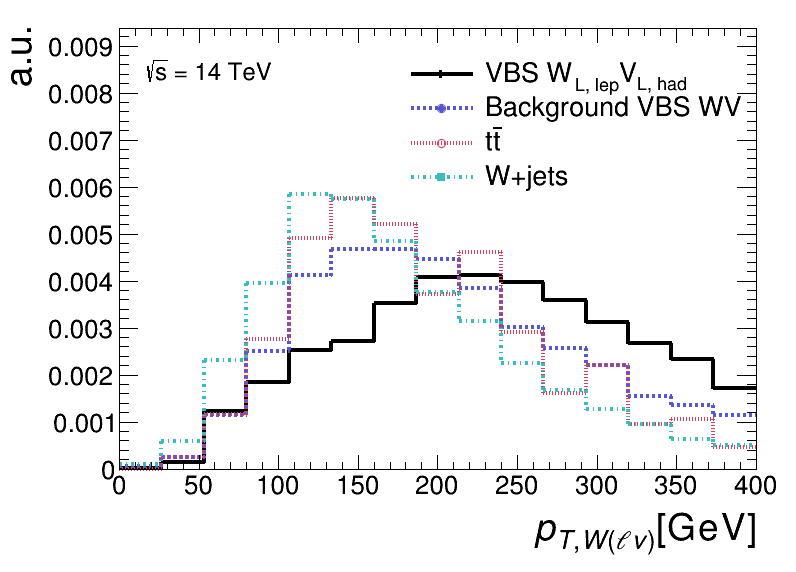}  
  \includegraphics[width=0.31\textwidth]{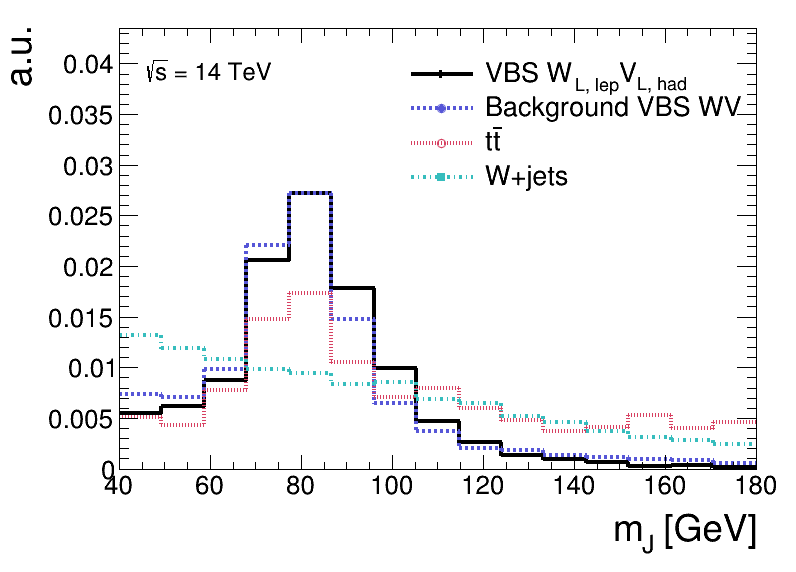} 
  \includegraphics[width=0.31\textwidth]{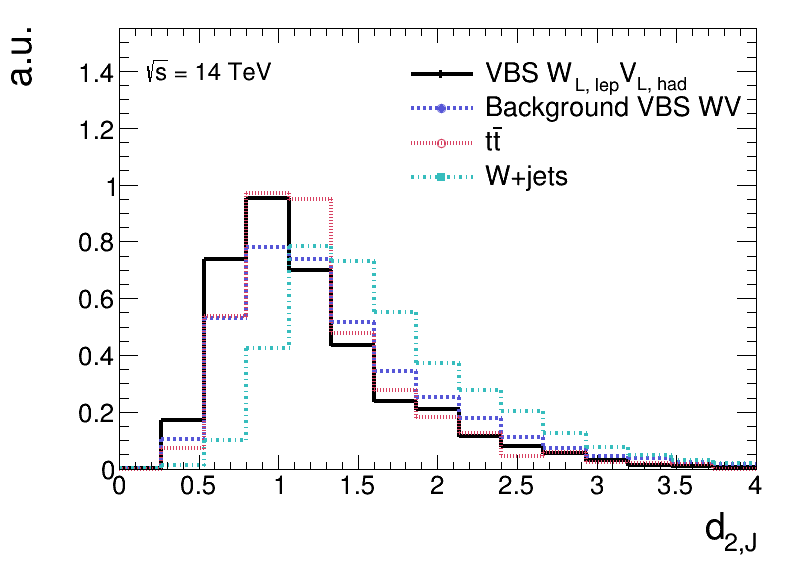} 
  \includegraphics[width=0.31\textwidth]{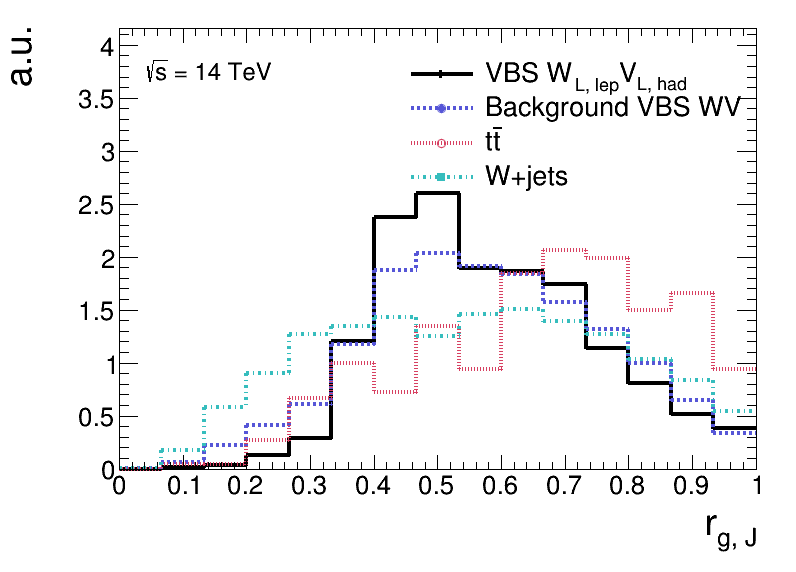} 
  \includegraphics[width=0.31\textwidth]{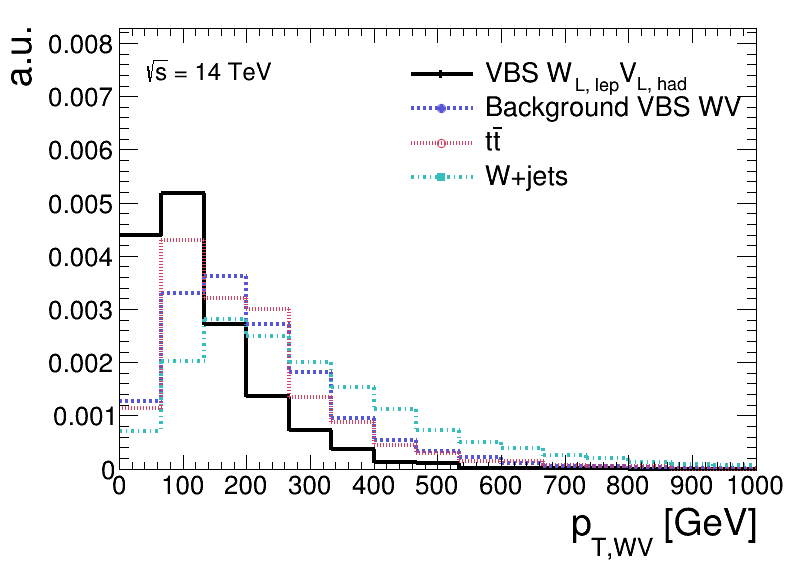} 
  \includegraphics[width=0.31\textwidth]{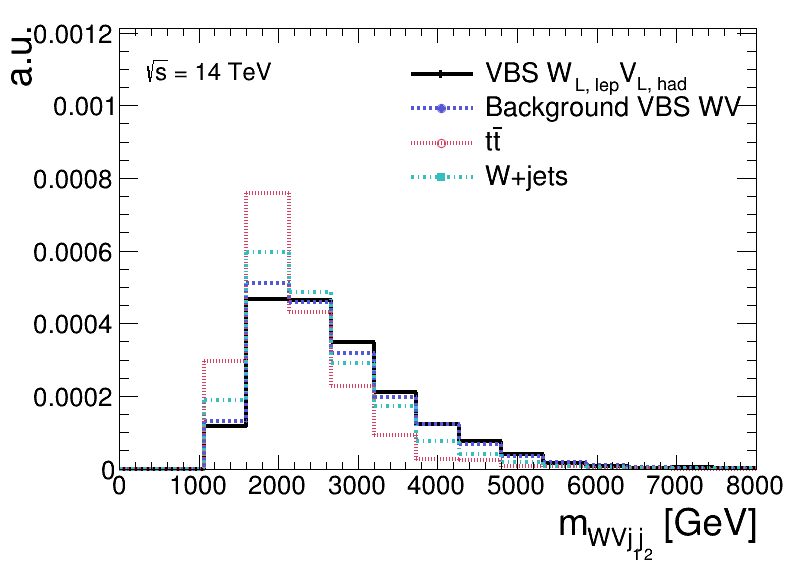}
  \includegraphics[width=0.31\textwidth]{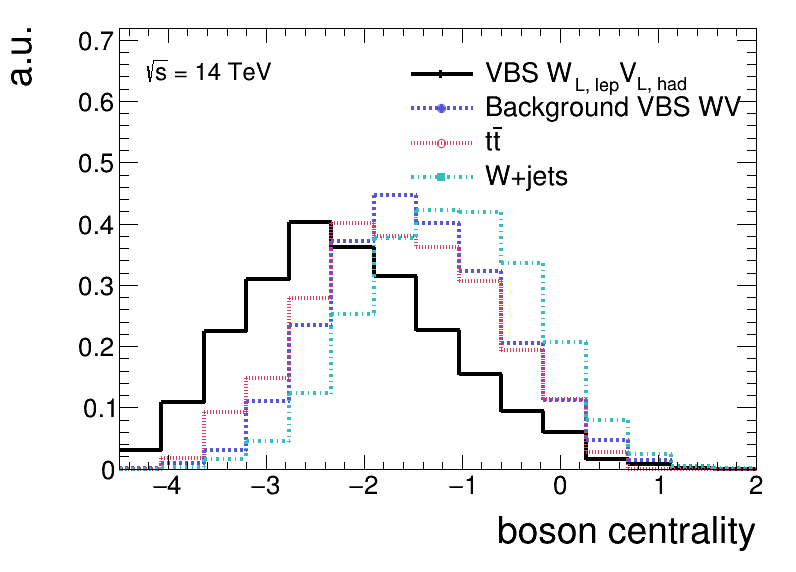} 
  \includegraphics[width=0.31\textwidth]{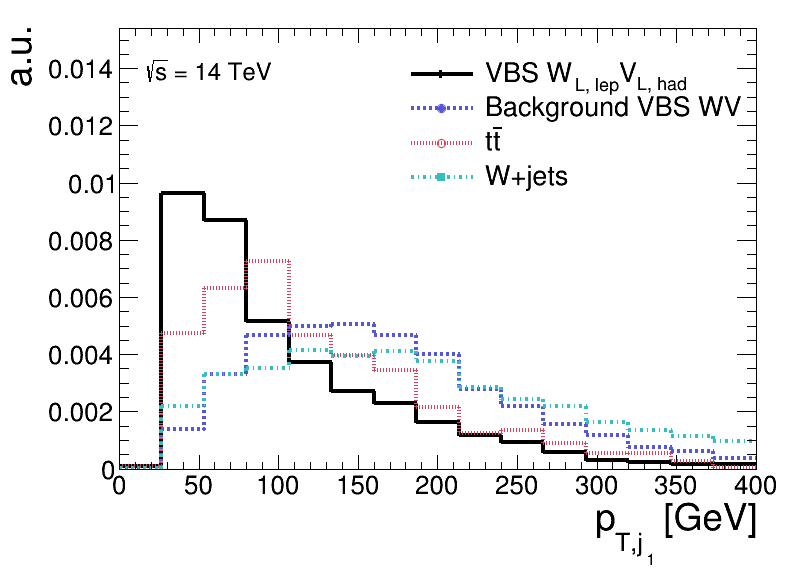} 
  \includegraphics[width=0.31\textwidth]{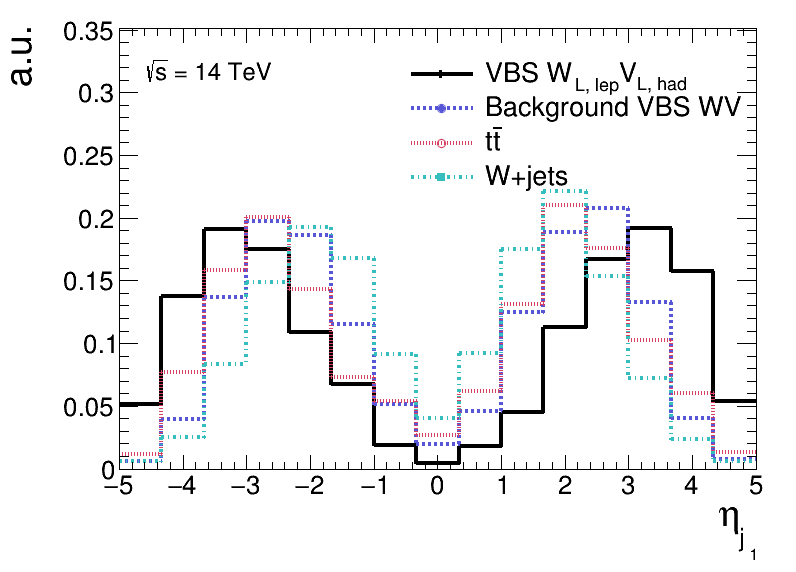} 
  \includegraphics[width=0.31\textwidth]{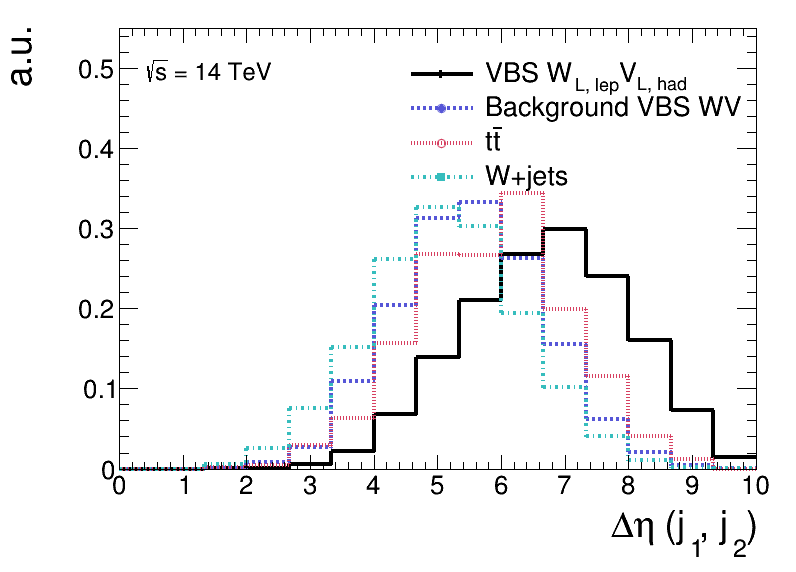} 
  \caption{Distributions of the different inputs to the multiclass tagger in $W+$jets, top-quark pair production, background VBS $WV$, and signal VBS $W_{L}V_{L}$ events.}
  \label{fig:taggerInputs}
\end{figure}

\begin{figure}[thb]
 \centering
 \includegraphics[width=0.45\textwidth]{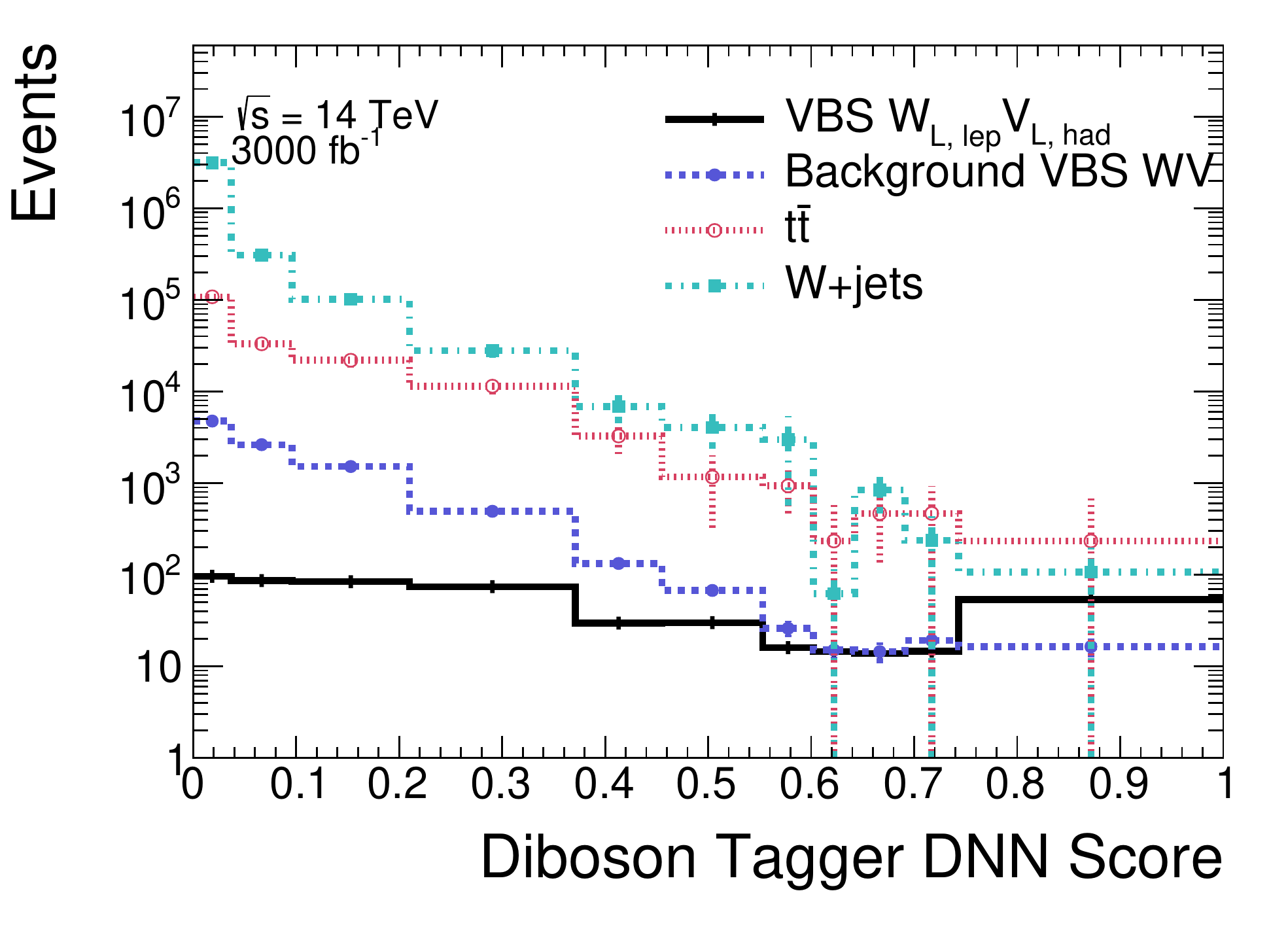}
 \includegraphics[width=0.45\textwidth]{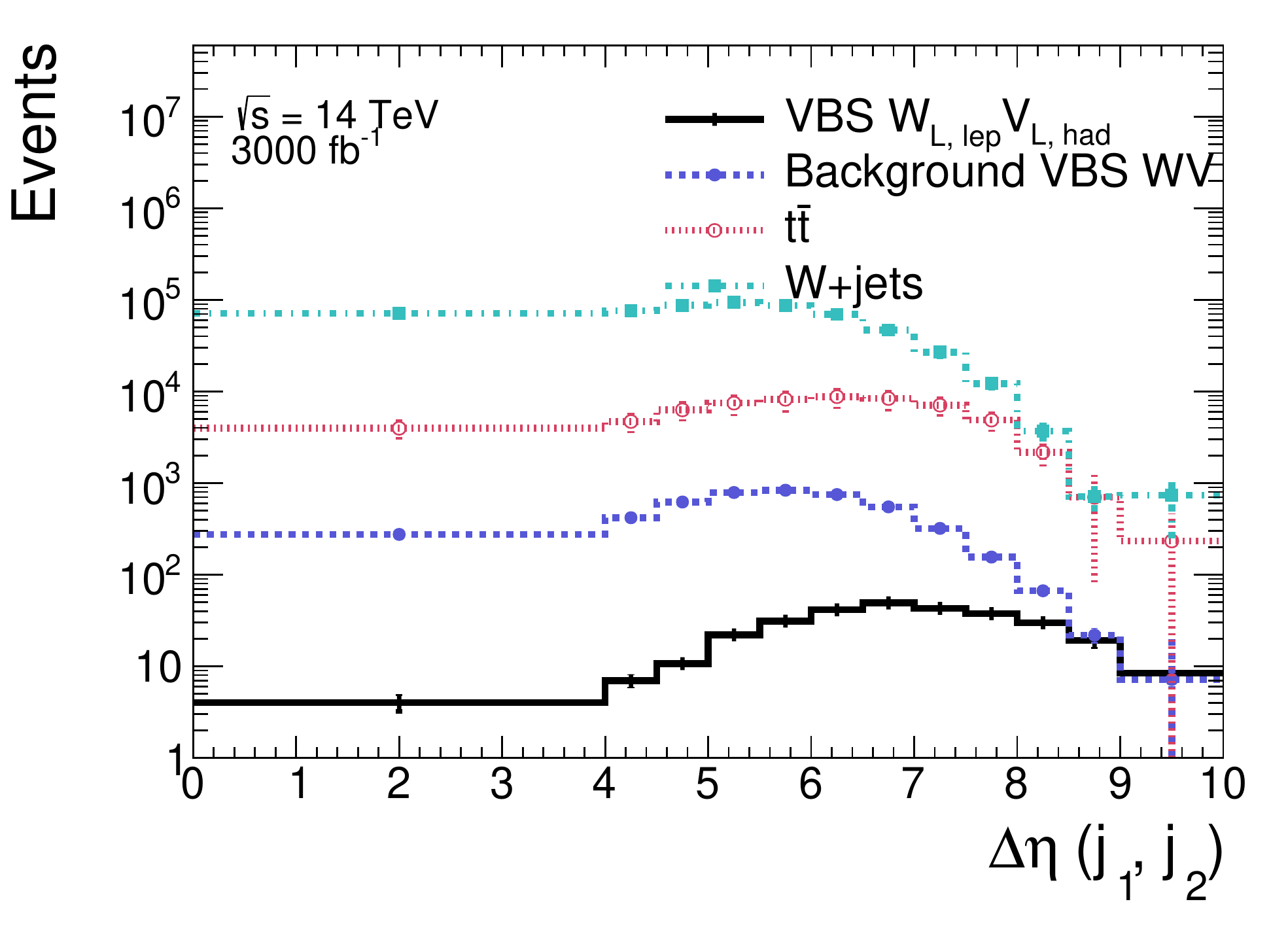}
 \caption{Signal fit input templates for (left) the multiclass DNN tagger, and (right) $\Delta \eta (j_{1}, j_{2})$.
          The error bars represent the statistical uncertainty.}
 \label{fig:taggerScore}
\end{figure}

The multiclass tagger is trained using the TMVA~\cite{Hoecker2007TMVAT} implementation of multiclass deep neural network (DNN) based on a multilayer perceptron with one hidden layer and 17 neurons.
Twelve variables, listed in Table~\ref{tab:taggerInputs}, are used as inputs into a multiclass DNN tagger.
\begin{table}[ht]
\centering
\caption{The variables used in the signal tagger.}
\label{tab:taggerInputs}
\begin{tabular}[t]{ll}
\hline
Variable & Description\\
\hline
$p_{T,\ell}$                     & $p_T$ of the charged lepton from the $W$ boson \\
$\eta_{\ell}$                     &  pseudorapidity of the charged lepton from the $W$ boson \\
$p_{T,W(\ell\nu)}$                  &  $p_T$ of the reconstructed leptonically decaying $W$ boson \\
$m_{J}$                          &  mass of the large-R jet\\
$d_{2,J}$                            &  ratio of three-point to two-point energy correlation functions\\
$r_{g,J}$                            &  angular separation between the two subjets\\
$p_{T, \mathrm{WV}}$                      &  $p_T$ of the diboson system\\
$m_{\mathrm{WVj_{1}j_{2}}}$                       &  mass of the $WVj_{1}j_{2}$ system\\
boson centrality                 &  min($\Delta \eta_{-}, \Delta \eta_{+}$), with\\
                                 &  $\Delta \eta_{-}$ = min[$\eta_{V_{had}}, \eta_{W_{lep}}$] - min[$\eta_{j_{1}}, \eta_{j_{2}}$]\\
                                 &  $\Delta \eta_{+}$ = max[$\eta_{V_{had}}, \eta_{W_{lep}}$] - max[$\eta_{j_{1}}, \eta_{j_{2}}$]\\
$p_{T,j_{1}}$                       &  $p_T$ of the leading tagging jet\\
$\eta_{j_{1}}$                      &  $\eta$ of the leading tagging jet\\
$\Delta \eta (j_{1}, j_{2})$             &  pseudorapidity difference between the two tagging jets\\
\hline
\end{tabular}
\end{table}
The distributions of these input variables are shown in Fig.~\ref{fig:taggerInputs} for both signal and background, with the pseudorapidity difference between the two tagging jets $\Delta \eta (j_{1}, j_{2})$ yielding the best single-variable signal discrimination.
For reference purposes, we compare signal extraction based on $\Delta \eta (j_{1}, j_{2})$ alone (while additionally requiring the jet mass to be 60~GeV$ < m_{J} < $100~GeV, and $d_{2,J} < 1.5$ to further reduce background contributions), with our DNN performance. The event yield for the DNN tagger score compared to $\Delta \eta (j_{1}, j_{2})$ is shown for the signal and background events in Fig.~\ref{fig:taggerScore}, illustrating that the discrimination power of the DNN score for the signal class is significantly better than the discrimination power of the most important input variable to the tagger $\Delta \eta (j_{1}, j_{2})$. This is expected, as the DNN is able to better separate the signal events from the background contributions by making full use of the kinematic information available in the event. The output of this DNN tagger for the signal class is used as an input to the template fit utilized to estimate the signal sensitivity, as described in the next section.

\section{Analysis sensitivity}
\label{sec:sensitivity}
The analysis sensitivity to the VBS $W_{L}V_{L}$ signal is extracted by performing a simultaneous binned maximum-likelihood fit to the signal and background distributions of the DNN and $\Delta \eta (j_{1}, j_{2})$, respectively.
A test statistic based on the profile likelihood ratio is used to test hypothesized values of the signal-strength factor.
The likelihood is defined as the product of the Poisson likelihood for each bin. 
The fit includes the main background contributions from W+jets and top-quark pair production, as well as the background contributions from the different polarization states.

While the sensitivity is limited by the statistical uncertainty, two sources of experimental uncertainties are considered: the jet energy and mass resolution of the large-R jet.
To evaluate these, the energy and mass of the jet are each smeared by 10\%. In addition, theoretical normalization uncertainties of 10\% are considered for each background. The normalization uncertainties are found to be dominant over the large-R jet energy and mass uncertainties.
Systematic uncertainties are taken into account as constrained nuisance parameters with Gaussian distributions.  
For each source of systematic uncertainty, the correlations across bins in the distributions and between different kinematic regions as well as those between signal and background are taken into account.

\begin{figure}[thb]
  \centering
  \includegraphics[width=0.60\textwidth]{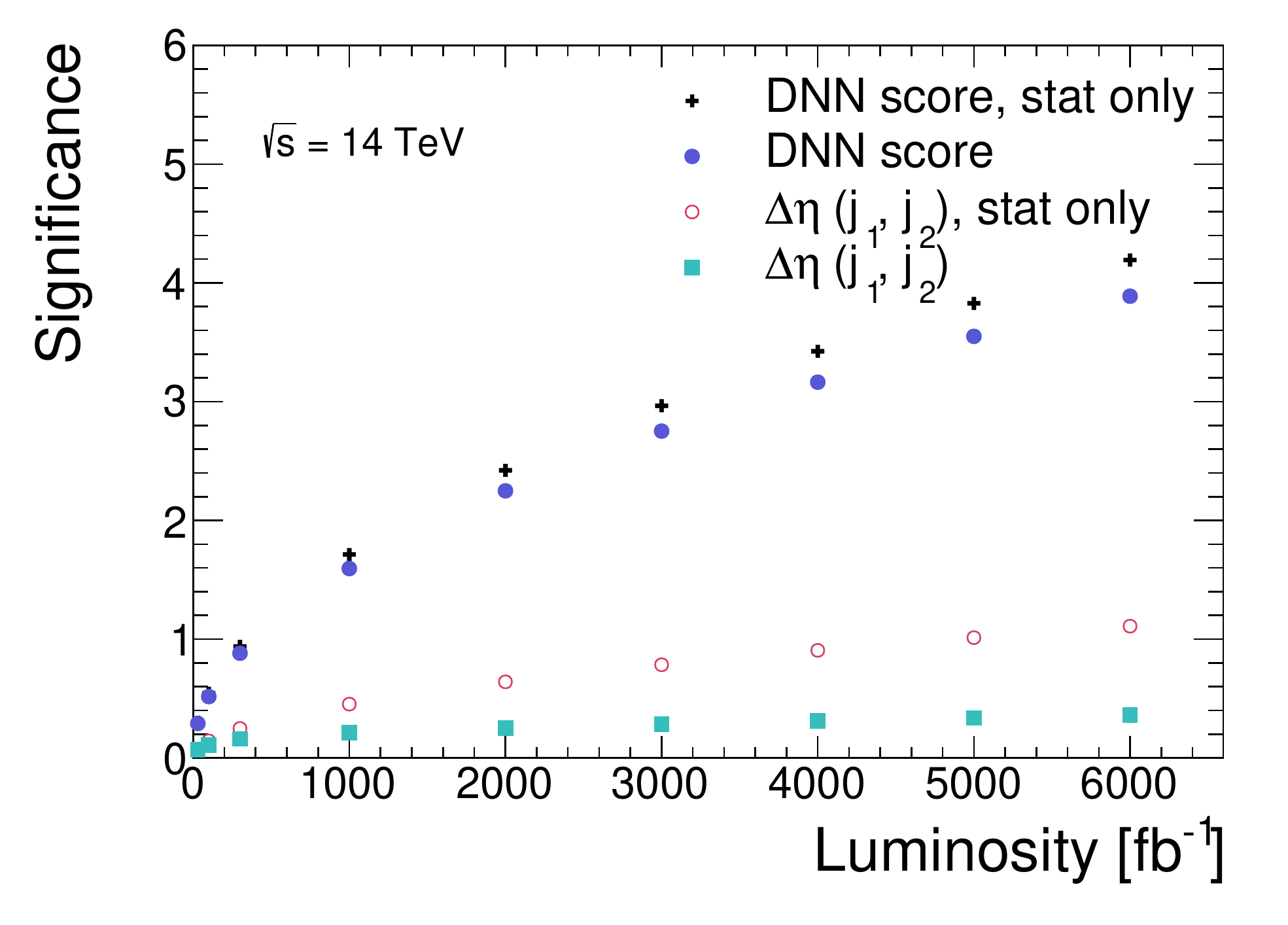}
  \caption{Expected $W_{L}V_{L}$ significance as a function of the integrated luminosity at the HL-LHC, using the full DNN and the $\Delta \eta (j_{1}, j_{2})$ observable alone. The ``stat only" results are obtained considering statistical uncertainties only, rather than both statistical and systematic uncertainties.}
  \label{fig:sensitivity}
\end{figure}

The expected significance is shown in Fig.~\ref{fig:sensitivity} as a function of the total integrated luminosity, with and without the inclusion of the systematic uncertainties.
The total integrated luminosity at the HL-LHC is expected to be 3000~fb$^{-1}$, and our results are shown for up to double this integrated
luminosity, giving a simple extrapolation to the expected sensitivity from the combination of measurements from ATLAS and CMS.
The sensitivity using the multiclass tagger is compared to using a $\Delta \eta (j_{1}, j_{2})$, which shows the best single-variable separation between signal and background.
The tagger provides significant gains over the single-variable input, demonstrating the importance of a multivariate tagger to improve the signal significance. 
The statistical uncertainties are the dominating factor, but some impact from the normalization uncertainties is seen.
With the expected luminosity of 3000~fb$^{-1}$ at the HL-LHC, the dataset may be used to separate the longitudinal component of VBS $WV$ with a significance of $3 \sigma$ considering statistical uncertainties only, and 2.8 $\sigma$ when including systematic uncertainties.

\section{Summary and Outlook}
\label{sec:summary}
The measurement of longitudinal VBS production is both a fundamental test of the SM and a window to new physics.
We have studied the prospects for measuring longitudinal VBS $WV$ production at the HL-LHC using the semileptonic final state 
where the $V$ boson hadronic decay products are boosted into a single large-radius jet.
Using substructure and machine learning techniques, our studies demonstrate that we can expect to establish longitudinal VBS production in this channel with approximately 3 standard deviations at the HL-LHC.
Despite the higher backgrounds and more complicated hadronic activity, this sensitivity is similar to what has been projected for the fully leptonic final states.
Further improvements may be achieved through the use of the resolved channel, $ZV jj$ semileptonic final states, as well as more complex object reconstruction. Even with the applied restrictions, our study demonstrates the importance of semileptonic final states in establishing longitudinal VBS production at the HL-LHC.

\section{Acknowledgements}
The work of V.C., M.-A.P. and J.R. is supported by the U.S. Department of Energy, Office of Science, Office of High Energy Physics under contract no. DE-SC0012704.

\clearpage
\bibliographystyle{JHEP}
\bibliography{polarization}

\providecommand{\href}[2]{#2}\begingroup\raggedright\begin{thebibliography}{10}

\bibitem{Dicus:1990fz}
D.~A. Dicus, J.~F. Gunion and R.~Vega, \emph{{Isolating the scattering of
  longitudinal W+'s at the SSC using like-sign dileptons}},
  \href{http://dx.doi.org/10.1016/0370-2693(91)91121-B}{\emph{Phys. Lett. B}
  {\bfseries 258} (1991) 475}.

\bibitem{Veltman:1976rt}
M.~J.~G. Veltman, \emph{{Second Threshold in Weak Interactions}}, {\emph{Acta
  Phys. Pol. B} {\bfseries 8} (1977) 475}.

\bibitem{Lee:1977yc}
B.~W. Lee, C.~Quigg and H.~B. Thacker, \emph{{Strength of Weak Interactions at
  Very High Energies and the Higgs Boson Mass}},
  \href{http://dx.doi.org/10.1103/PhysRevLett.38.883}{\emph{Phys. Rev. Lett.}
  {\bfseries 38} (1977) 883}.

\bibitem{Lee:1977eg}
B.~W. Lee, C.~Quigg and H.~B. Thacker, \emph{{Weak interactions at very high
  energies: The role of the Higgs-boson mass}},
  \href{http://dx.doi.org/10.1103/PhysRevD.16.1519}{\emph{Phys. Rev. D}
  {\bfseries 16} (1977) 1519}.

\bibitem{Gunion:1989we}
J.~F. Gunion, H.~E. Haber, G.~L. Kane and S.~Dawson, \emph{{The Higgs Hunter's
  Guide}}, {\emph{Front. Phys.} {\bfseries 80} (2000) }.

\bibitem{Aad:2012tfa}
{\scshape ATLAS} collaboration, G.~Aad et~al., \emph{{Observation of a new
  particle in the search for the Standard Model Higgs boson with the ATLAS
  detector at the LHC}},
  \href{http://dx.doi.org/10.1016/j.physletb.2012.08.020}{\emph{Phys. Lett. B}
  {\bfseries 716} (2012) 1--29},
  [\href{https://arxiv.org/abs/1207.7214}{{\ttfamily 1207.7214}}].

\bibitem{Chatrchyan:2012ufa}
{\scshape CMS} collaboration, S.~Chatrchyan et~al., \emph{{Observation of a new
  boson at a mass of 125 GeV with the CMS experiment at the LHC}},
  \href{http://dx.doi.org/10.1016/j.physletb.2012.08.021}{\emph{Phys. Lett. B}
  {\bfseries 716} (2012) 30--61},
  [\href{https://arxiv.org/abs/1207.7235}{{\ttfamily 1207.7235}}].

\bibitem{Khachatryan:2014kca}
{\scshape CMS} collaboration, V.~Khachatryan et~al., \emph{{Constraints on the
  spin-parity and anomalous HVV couplings of the Higgs boson in proton
  collisions at 7 and 8 TeV}},
  \href{http://dx.doi.org/10.1103/PhysRevD.92.012004}{\emph{Phys. Rev. D}
  {\bfseries 92} (2015) 012004},
  [\href{https://arxiv.org/abs/1411.3441}{{\ttfamily 1411.3441}}].

\bibitem{Khachatryan:2014jba}
{\scshape CMS} collaboration, V.~Khachatryan et~al., \emph{{Precise
  determination of the mass of the Higgs boson and tests of compatibility of
  its couplings with the standard model predictions using proton collisions at
  7 and 8 $\,\text {TeV}$}},
  \href{http://dx.doi.org/10.1140/epjc/s10052-015-3351-7}{\emph{Eur. Phys. J.
  C} {\bfseries 75} (2015) 212},
  [\href{https://arxiv.org/abs/1412.8662}{{\ttfamily 1412.8662}}].

\bibitem{Aad:2015mxa}
{\scshape ATLAS} collaboration, G.~Aad et~al., \emph{{Study of the spin and
  parity of the Higgs boson in diboson decays with the ATLAS detector}},
  \href{http://dx.doi.org/10.1140/epjc/s10052-015-3685-1}{\emph{Eur. Phys. J.
  C} {\bfseries 75} (2015) 476},
  [\href{https://arxiv.org/abs/1506.05669}{{\ttfamily 1506.05669}}].

\bibitem{Aad:2015gba}
{\scshape ATLAS} collaboration, G.~Aad et~al., \emph{{Measurements of the Higgs
  boson production and decay rates and coupling strengths using pp collision
  data at $\sqrt{s}=7$ and 8 TeV in the ATLAS experiment}},
  \href{http://dx.doi.org/10.1140/epjc/s10052-015-3769-y}{\emph{Eur. Phys. J.
  C} {\bfseries 76} (2016) 6},
  [\href{https://arxiv.org/abs/1507.04548}{{\ttfamily 1507.04548}}].

\bibitem{Eboli:2006wa}
O.~J.~P. Eboli, M.~C. Gonzalez-Garcia and J.~K. Mizukoshi, \emph{{p p
  ---\ensuremath{>} $j j e^{\pm} \mu^\pm \nu \nu$ and $j j e^{\pm} \mu^\mp \nu
  \nu$ at O( $\alpha_{em}^6$) and O($\alpha_{em}^4 \alpha_{s}^2$) for the study
  of the quartic electroweak gauge boson vertex at CERN LHC}},
  \href{http://dx.doi.org/10.1103/PhysRevD.74.073005}{\emph{Phys. Rev. D}
  {\bfseries 74} (2006) 073005},
  [\href{https://arxiv.org/abs/hep-ph/0606118}{{\ttfamily hep-ph/0606118}}].

\bibitem{Campbell:2015vwa}
J.~M. Campbell and R.~K. Ellis, \emph{{Higgs constraints from vector boson
  fusion and scattering}},
  \href{http://dx.doi.org/10.1007/JHEP04(2015)030}{\emph{J. High Energy Phys.}
  {\bfseries 04} (2015) 030},
  [\href{https://arxiv.org/abs/1502.02990}{{\ttfamily 1502.02990}}].

\bibitem{Alboteanu:2008my}
A.~Alboteanu, W.~Kilian and J.~Reuter, \emph{{Resonances and unitarity in weak
  boson scattering at the LHC}},
  \href{http://dx.doi.org/10.1088/1126-6708/2008/11/010}{\emph{J. High Energy
  Phys.} {\bfseries 11} (2008) 010},
  [\href{https://arxiv.org/abs/0806.4145}{{\ttfamily 0806.4145}}].

\bibitem{Godfrey:2010qb}
S.~Godfrey and K.~Moats, \emph{{Exploring Higgs triplet models via vector boson
  scattering at the LHC}},
  \href{http://dx.doi.org/10.1103/PhysRevD.81.075026}{\emph{Phys. Rev. D}
  {\bfseries 81} (2010) 075026},
  [\href{https://arxiv.org/abs/1003.3033}{{\ttfamily 1003.3033}}].

\bibitem{Espriu:2012ih}
D.~Espriu and B.~Yencho, \emph{{Longitudinal WW scattering in light of the
  \textquotedblleft{}Higgs boson\textquotedblright{} discovery}},
  \href{http://dx.doi.org/10.1103/PhysRevD.87.055017}{\emph{Phys. Rev. D}
  {\bfseries 87} (2013) 055017},
  [\href{https://arxiv.org/abs/1212.4158}{{\ttfamily 1212.4158}}].

\bibitem{Chang:2013aya}
J.~Chang, K.~Cheung, C.-T. Lu and T.-C. Yuan, \emph{{WW scattering in the era
  of post-Higgs-boson discovery}},
  \href{http://dx.doi.org/10.1103/PhysRevD.87.093005}{\emph{Phys. Rev. D}
  {\bfseries 87} (2013) 093005},
  [\href{https://arxiv.org/abs/1303.6335}{{\ttfamily 1303.6335}}].

\bibitem{Chiang:2014bia}
C.-W. Chiang, S.~Kanemura and K.~Yagyu, \emph{{Novel constraint on the
  parameter space of the Georgi-Machacek model with current LHC data}},
  \href{http://dx.doi.org/10.1103/PhysRevD.90.115025}{\emph{Phys. Rev. D}
  {\bfseries 90} (2014) 115025},
  [\href{https://arxiv.org/abs/1407.5053}{{\ttfamily 1407.5053}}].

\bibitem{Kilian:2014zja}
W.~Kilian, T.~Ohl, J.~Reuter and M.~Sekulla, \emph{{High-energy vector boson
  scattering after the Higgs boson discovery}},
  \href{http://dx.doi.org/10.1103/PhysRevD.91.096007}{\emph{Phys. Rev. D}
  {\bfseries 91} (2015) 096007},
  [\href{https://arxiv.org/abs/1408.6207}{{\ttfamily 1408.6207}}].

\bibitem{Aaboud:2018ddq}
{\scshape ATLAS} collaboration, M.~Aaboud et~al., \emph{{Observation of
  electroweak $W^{\pm}Z$ boson pair production in association with two jets in
  $pp$ collisions at $\sqrt{s} =$ 13 TeV with the ATLAS detector}},
  \href{http://dx.doi.org/10.1016/j.physletb.2019.05.012}{\emph{Phys. Lett. B}
  {\bfseries 793} (2019) 469--492},
  [\href{https://arxiv.org/abs/1812.09740}{{\ttfamily 1812.09740}}].

\bibitem{Sirunyan:2020gyx}
{\scshape CMS} collaboration, A.~M. Sirunyan et~al., \emph{{Measurements of
  production cross sections of WZ and same-sign WW boson pairs in association
  with two jets in proton-proton collisions at $\sqrt{s} =$ 13 TeV}},
  \href{http://dx.doi.org/10.1016/j.physletb.2020.135710}{\emph{Phys. Lett. B}
  {\bfseries 809} (2020) 135710},
  [\href{https://arxiv.org/abs/2005.01173}{{\ttfamily 2005.01173}}].

\bibitem{Aaboud:2019nmv}
{\scshape ATLAS} collaboration, M.~Aaboud et~al., \emph{{Observation of
  Electroweak Production of a Same-Sign $W$ Boson Pair in Association with Two
  Jets in $pp$ Collisions at $\sqrt{s}=13$ TeV with the ATLAS Detector}},
  \href{http://dx.doi.org/10.1103/PhysRevLett.123.161801}{\emph{Phys. Rev.
  Lett.} {\bfseries 123} (2019) 161801},
  [\href{https://arxiv.org/abs/1906.03203}{{\ttfamily 1906.03203}}].

\bibitem{Sirunyan:2017ret}
{\scshape CMS} collaboration, A.~M. Sirunyan et~al., \emph{{Observation of
  Electroweak Production of Same-Sign W Boson Pairs in the Two Jet and Two
  Same-Sign Lepton Final State in Proton-Proton Collisions at $\sqrt{s} = $ 13
  TeV}}, \href{http://dx.doi.org/10.1103/PhysRevLett.120.081801}{\emph{Phys.
  Rev. Lett.} {\bfseries 120} (2018) 081801},
  [\href{https://arxiv.org/abs/1709.05822}{{\ttfamily 1709.05822}}].

\bibitem{Aad:2020zbq}
{\scshape ATLAS} collaboration, G.~Aad et~al., \emph{{Observation of
  Electroweak Production of Two Jets and a $Z$-Boson Pair with the ATLAS
  detector at the LHC}},  \href{https://arxiv.org/abs/2004.10612}{{\ttfamily
  2004.10612}}.

\bibitem{Sirunyan:2020alo}
{\scshape CMS} collaboration, A.~M. Sirunyan et~al., \emph{{Evidence for
  electroweak production of four charged leptons and two jets in proton-proton
  collisions at $\sqrt {s}$ = 13 TeV}},
  \href{http://dx.doi.org/10.1016/j.physletb.2020.135992}{\emph{Phys. Lett. B}
  {\bfseries 812} (2021) 135992},
  [\href{https://arxiv.org/abs/2008.07013}{{\ttfamily 2008.07013}}].

\bibitem{Aaboud:2016uuk}
{\scshape ATLAS} collaboration, M.~Aaboud et~al., \emph{{Search for anomalous
  electroweak production of $WW/WZ$ in association with a high-mass dijet
  system in $pp$ collisions at $\sqrt{s}=8$ TeV with the ATLAS detector}},
  \href{http://dx.doi.org/10.1103/PhysRevD.95.032001}{\emph{Phys. Rev. D}
  {\bfseries 95} (2017) 032001},
  [\href{https://arxiv.org/abs/1609.05122}{{\ttfamily 1609.05122}}].

\bibitem{Sirunyan:2019der}
{\scshape CMS} collaboration, A.~M. Sirunyan et~al., \emph{{Search for
  anomalous electroweak production of vector boson pairs in association with
  two jets in proton-proton collisions at 13 TeV}},
  \href{http://dx.doi.org/10.1016/j.physletb.2019.134985}{\emph{Phys. Lett. B}
  {\bfseries 798} (2019) 134985},
  [\href{https://arxiv.org/abs/1905.07445}{{\ttfamily 1905.07445}}].

\bibitem{Aad:2019xxo}
{\scshape ATLAS} collaboration, G.~Aad et~al., \emph{{Search for electroweak
  diboson production in association with a high-mass dijet system in
  semileptonic final states in $pp$ collisions at $\sqrt{s}=13$ TeV with the
  ATLAS detector}},
  \href{http://dx.doi.org/10.1103/PhysRevD.100.032007}{\emph{Phys. Rev. D}
  {\bfseries 100} (2019) 032007},
  [\href{https://arxiv.org/abs/1905.07714}{{\ttfamily 1905.07714}}].

\bibitem{ATLAS:2018ocj}
{\scshape ATLAS} collaboration, \emph{{HL-LHC prospects for diboson resonance
  searches and electroweak vector boson scattering in the $WW/WZ\to\ell\nu qq$
  final state}},  ATL-PHYS-PUB-2018-022, 2018,
  \url{https://inspirehep.net/literature/1795287}.

\bibitem{Aaboud:2019gxl}
{\scshape ATLAS} collaboration, M.~Aaboud et~al., \emph{{Measurement of
  $W^{\pm}Z$ production cross sections and gauge boson polarisation in $pp$
  collisions at $\sqrt{s} = 13$ TeV with the ATLAS detector}},
  \href{http://dx.doi.org/10.1140/epjc/s10052-019-7027-6}{\emph{Eur. Phys. J.
  C} {\bfseries 79} (2019) 535},
  [\href{https://arxiv.org/abs/1902.05759}{{\ttfamily 1902.05759}}].

\bibitem{Sirunyan:2020gvn}
{\scshape CMS} collaboration, A.~M. Sirunyan et~al., \emph{{Measurements of
  production cross sections of polarized same-sign W boson pairs in association
  with two jets in proton-proton collisions at $\sqrt{s} =$ 13 TeV}},
  \href{http://dx.doi.org/10.1016/j.physletb.2020.136018}{\emph{Phys. Lett. B}
  {\bfseries 812} (2021) 136018},
  [\href{https://arxiv.org/abs/2009.09429}{{\ttfamily 2009.09429}}].

\bibitem{CMSCollaboration:2015zni}
\emph{{Technical Proposal for the Phase-II Upgrade of the CMS Detector}},
  CERN-LHCC-2015-010, LHCC-P-008, CMS-TDR-15-02, 6, 2015,
  \url{https://inspirehep.net/literature/1614097}.

\bibitem{CMS:2018mbt}
{\scshape CMS} collaboration, \emph{{Vector Boson Scattering prospective
  studies in the ZZ fully leptonic decay channel for the High-Luminosity and
  High-Energy LHC upgrades}},  CMS-PAS-FTR-18-014, 2018,
  \url{https://inspirehep.net/literature/1708645}.

\bibitem{CMS:2018ylh}
{\scshape CMS} collaboration, \emph{{Prospects for the measurement of
  electroweak and polarized WZ to 3lv production cross sections at the
  High-Luminosity LHC}},  CMS-PAS-FTR-18-038, 2018,
  \url{https://inspirehep.net/literature/1708642}.

\bibitem{ATLAS:2018tav}
{\scshape ATLAS} collaboration, \emph{{Prospective study of vector boson
  scattering in WZ fully leptonic final state at HL-LHC}},
  ATL-PHYS-PUB-2018-023, 2018, \url{https://inspirehep.net/literature/1795274}.

\bibitem{CMS:2018zxa}
{\scshape CMS} collaboration, \emph{{Study of W$^\pm$W$^\pm$ production via
  vector boson scattering at the HL-LHC with the upgraded CMS detector}},
  CMS-PAS-FTR-18-005, 2018, \url{https://inspirehep.net/literature/1703668}.

\bibitem{ATLAS:2018uld}
{\scshape ATLAS} collaboration, \emph{{Prospects for the measurement of the
  $W^{\pm}W^{\pm}$ scattering cross section and extraction of the longitudinal
  scattering component in $pp$ collisions at the High-Luminosity LHC with the
  ATLAS experiment}},  ATL-PHYS-PUB-2018-052, 2018,
  \url{https://inspirehep.net/literature/1795250}.

\bibitem{Searcy:2015apa}
J.~Searcy, L.~Huang, M.-A. Pleier and J.~Zhu, \emph{{Determination of the $WW$
  polarization fractions in $pp \to W^\pm W^\pm jj$ using a deep machine
  learning technique}},
  \href{http://dx.doi.org/10.1103/PhysRevD.93.094033}{\emph{Phys. Rev. D}
  {\bfseries 93} (2016) 094033},
  [\href{https://arxiv.org/abs/1510.01691}{{\ttfamily 1510.01691}}].

\bibitem{Grossi:2020orx}
M.~Grossi, J.~Novak, B.~Kersevan and D.~Rebuzzi, \emph{{Comparing traditional
  and deep-learning techniques of kinematic reconstruction for polarization
  discrimination in vector boson scattering}},
  \href{http://dx.doi.org/10.1140/epjc/s10052-020-08713-1}{\emph{Eur. Phys. J.
  C} {\bfseries 80} (2020) 1144},
  [\href{https://arxiv.org/abs/2008.05316}{{\ttfamily 2008.05316}}].

\bibitem{Cavaliere:2018zcf}
V.~Cavaliere, R.~Les, T.~Nitta and K.~Terashi, \emph{{HE-LHC prospects for
  diboson resonance searches and electroweak WW/WZ production via vector boson
  scattering in the semi-leptonic final states}},
  \href{https://arxiv.org/abs/1812.00841}{{\ttfamily 1812.00841}}.

\bibitem{Alwall:2014hca}
J.~Alwall, R.~Frederix, S.~Frixione, V.~Hirschi, F.~Maltoni, O.~Mattelaer
  et~al., \emph{{The automated computation of tree-level and next-to-leading
  order differential cross sections, and their matching to parton shower
  simulations}}, \href{http://dx.doi.org/10.1007/JHEP07(2014)079}{\emph{J. High
  Energy Phys.} {\bfseries 07} (2014) 079},
  [\href{https://arxiv.org/abs/1405.0301}{{\ttfamily 1405.0301}}].

\bibitem{Sjostrand:2007gs}
T.~Sjostrand, S.~Mrenna and P.~Z. Skands, \emph{{A brief introduction to PYTHIA
  8.1}}, \href{http://dx.doi.org/10.1016/j.cpc.2008.01.036}{\emph{Comput. Phys.
  Commun.} {\bfseries 178} (2008) 852--867},
  [\href{https://arxiv.org/abs/0710.3820}{{\ttfamily 0710.3820}}].

\bibitem{BuarqueFranzosi:2019boy}
D.~Buarque~Franzosi, O.~Mattelaer, R.~Ruiz and S.~Shil, \emph{{Automated
  predictions from polarized matrix elements}},
  \href{http://dx.doi.org/10.1007/JHEP04(2020)082}{\emph{J. High Energy Phys.}
  {\bfseries 04} (2020) 082},
  [\href{https://arxiv.org/abs/1912.01725}{{\ttfamily 1912.01725}}].

\bibitem{Catani:2001cc}
S.~Catani, F.~Krauss, R.~Kuhn and B.~R. Webber, \emph{{QCD Matrix Elements +
  Parton Showers}},
  \href{http://dx.doi.org/10.1088/1126-6708/2001/11/063}{\emph{J. High Energy
  Phys.} {\bfseries 11} (2001) 063},
  [\href{https://arxiv.org/abs/hep-ph/0109231}{{\ttfamily hep-ph/0109231}}].

\bibitem{Lonnblad:2001iq}
L.~Lonnblad, \emph{{Correcting the Colour-Dipole Cascade Model with Fixed Order
  Matrix Elements}},
  \href{http://dx.doi.org/10.1088/1126-6708/2002/05/046}{\emph{J. High Energy
  Phys.} {\bfseries 05} (2002) 046},
  [\href{https://arxiv.org/abs/hep-ph/0112284}{{\ttfamily hep-ph/0112284}}].

\bibitem{Artoisenet:2012st}
P.~Artoisenet, R.~Frederix, O.~Mattelaer and R.~Rietkerk, \emph{{Automatic
  spin-entangled decays of heavy resonances in Monte Carlo simulations}},
  \href{http://dx.doi.org/10.1007/JHEP03(2013)015}{\emph{J. High Energy Phys.}
  {\bfseries 03} (2013) 015},
  [\href{https://arxiv.org/abs/1212.3460}{{\ttfamily 1212.3460}}].

\bibitem{deFavereau:2013fsa}
{\scshape DELPHES 3} collaboration, J.~de~Favereau, C.~Delaere, P.~Demin,
  A.~Giammanco, V.~Lema\^\i{}tre, A.~Mertens et~al., \emph{{DELPHES 3: a
  modular framework for fast simulation of a generic collider experiment}},
  \href{http://dx.doi.org/10.1007/JHEP02(2014)057}{\emph{J. High Energy Phys.}
  {\bfseries 02} (2014) 057},
  [\href{https://arxiv.org/abs/1307.6346}{{\ttfamily 1307.6346}}].

\bibitem{Azzi:2019yne}
P.~Azzi et~al., \emph{{Report from Working Group 1}: {Standard Model physics at
  the HL-LHC and HE-LHC}},
  \href{http://dx.doi.org/10.23731/CYRM-2019-007.1}{\emph{CERN Yellow Rep.
  Monogr.} {\bfseries 7} (2019) 1--220},
  [\href{https://arxiv.org/abs/1902.04070}{{\ttfamily 1902.04070}}].

\bibitem{Atlas:2019qfx}
{\scshape ATLAS, CMS} collaboration, \emph{{Addendum to the report on the
  physics at the HL-LHC, and perspectives for the HE-LHC: Collection of notes
  from ATLAS and CMS}},
  \href{http://dx.doi.org/10.23731/CYRM-2019-007.Addendum}{\emph{CERN Yellow
  Rep. Monogr.} {\bfseries 7} (2019) Addendum},
  [\href{https://arxiv.org/abs/1902.10229}{{\ttfamily 1902.10229}}].

\bibitem{Cacciari:2011ma}
M.~Cacciari, G.~P. Salam and G.~Soyez, \emph{{FastJet User Manual}},
  \href{http://dx.doi.org/10.1140/epjc/s10052-012-1896-2}{\emph{Eur. Phys. J.
  C} {\bfseries 72} (2012) 1896},
  [\href{https://arxiv.org/abs/1111.6097}{{\ttfamily 1111.6097}}].

\bibitem{Cacciari:2008gp}
M.~Cacciari, G.~P. Salam and G.~Soyez, \emph{{The anti-$k_t$ jet clustering
  algorithm}}, \href{http://dx.doi.org/10.1088/1126-6708/2008/04/063}{\emph{J.
  High Energy Phys.} {\bfseries 04} (2008) 063},
  [\href{https://arxiv.org/abs/0802.1189}{{\ttfamily 0802.1189}}].

\bibitem{Larkoski:2014wba}
A.~J. Larkoski, S.~Marzani, G.~Soyez and J.~Thaler, \emph{{Soft drop}},
  \href{http://dx.doi.org/10.1007/JHEP05(2014)146}{\emph{J. High Energy Phys.}
  {\bfseries 05} (2014) 146},
  [\href{https://arxiv.org/abs/1402.2657}{{\ttfamily 1402.2657}}].

\bibitem{Larkoski:2015kga}
A.~J. Larkoski, I.~Moult and D.~Neill, \emph{{Analytic boosted boson
  discrimination}}, \href{http://dx.doi.org/10.1007/JHEP05(2016)117}{\emph{J.
  High Energy Phys.} {\bfseries 05} (2016) 117},
  [\href{https://arxiv.org/abs/1507.03018}{{\ttfamily 1507.03018}}].

\bibitem{Hoecker2007TMVAT}
A.~Hoecker, P.~Speckmayer, J.~Stelzer, J.~Therhaag, E.~V. Toerne, H.~Voss
  et~al., \emph{{TMVA - Toolkit for Multivariate Data Analysis}},  2007,
  \url{https://inspirehep.net/literature/746087}.

\end{thebibliography}\endgroup

\end{document}